\documentclass[a4paper,11pt]{article}
\pdfoutput=1 

\usepackage{jcappub} 

\usepackage[T1]{fontenc} 

\usepackage{graphicx}
\usepackage{latexsym}
\usepackage{amsfonts}
\usepackage{amssymb}
\usepackage{amsmath}
\usepackage{slashed}
\usepackage{mathtools}
\usepackage{hyperref}
\usepackage{url}
\usepackage{color}
\usepackage{cancel}
\usepackage{multirow,array}
\usepackage{float}
\usepackage{subfigure}
\usepackage{dcolumn} 
\usepackage{booktabs}

\usepackage[normalem]{ulem} 

\usepackage{tikz}
\usetikzlibrary{positioning}
\usetikzlibrary{calc}
\usetikzlibrary{backgrounds}
\makeatletter
\newcommand\footnoteref[1]{\protected@xdef\@thefnmark{\ref{#1}}\@footnotemark}
\makeatother

\newcommand{\w}{\phantom{0}} 
\newcommand{\Gaia}{\textit{Gaia}}

\newcommand{\localdensity}{1.23} 
\newcommand{\localdensityerr}{0.13} 
\newcommand{\localdensityGeV}{0.47} 
\newcommand{\localdensityerrGeV}{0.05} 

\newcommand{\aX}{-1.93} 
\newcommand{\aXerr}{0.08} 
\newcommand{\aY}{0.08} 
\newcommand{\aYerr}{0.05} 
\newcommand{\aZ}{-0.05} 
\newcommand{\aZerr}{0.06} 
\newcommand{\aMag}{1.93} 
\newcommand{\aMagerr}{0.08} 
\newcommand{\aR}{-1.93} 
\newcommand{\aRerr}{0.08} 

\title{\boldmath Mapping Dark Matter in the Milky Way using Normalizing Flows and Gaia DR3}

\author[a,b]{Sung Hak Lim}
\author[b]{Eric Putney,}
\author[b]{Matthew R.~Buckley,}
\author[b]{and David Shih}

\affiliation[a]{Particle Theory and Cosmology Group, Center for Theoretical Physics of the Universe, \\Institute for Basic Science (IBS), \\55 Expo-ro, Yuseong-gu, Daejeon 34126, Republic of Korea}
\affiliation[b]{NHETC, Department of Physics and Astronomy,\\Rutgers, the State University of New Jersey,\\126 Frelinghuysen Road, Piscataway, NJ 08854, USA}

\emailAdd{sunghak.lim@rutgers.edu}
\emailAdd{eputney@physics.rutgers.edu}
\emailAdd{mbuckley@physics.rutgers.edu}
\emailAdd{shih@physics.rutgers.edu}

\abstract{
We present a novel, data-driven analysis of Galactic dynamics, using unsupervised machine learning -- in the form of density estimation with normalizing flows -- to learn the underlying phase space distribution of 6 million nearby stars from the \Gaia{} DR3 catalog. 
Solving the equilibrium collisionless Boltzmann equation, we calculate -- for the first time ever -- a model-free, unbinned estimate of the local acceleration and mass density fields within a 3~kpc sphere around the Sun. 
As our approach makes no assumptions about symmetries, we can test for signs of disequilibrium in our results. 
We find our results are consistent with equilibrium at the 10\% level, limited by the current precision of the normalizing flows. 
After subtracting the known contribution of stars and gas from the calculated mass density, we find clear evidence for dark matter throughout the analyzed volume. 
Assuming spherical symmetry and averaging mass density measurements, we find a local dark matter density of $\localdensityGeV\pm \localdensityerrGeV$~GeV/cm$^3$. 
We compute the dark matter density at four radii in the stellar halo and fit to a generalized NFW profile. Although the uncertainties are large, we find a profile broadly consistent with recent analyses.
}

\arxivnumber{2305.13358}

\usepackage{fancyhdr}
\fancypagestyle{titlepage}{
  \fancyhf{}
  \fancyhead[R]{\raggedleft CTPU-PTC-25-01}
}
\fancypagestyle{otherpage}{
    \fancyhf{}
    \fancyhead[R]{}
    \fancyhead[L]{}
    \fancyhead[C]{}
}

\begin{document}
\maketitle
\flushbottom

\section{Introduction}
Multiple lines of evidence indicate that the majority of matter in the Universe is dark -- that is, it does not interact with the known particles through electromagnetic or strong nuclear interactions. Measurements of galaxy rotation curves \cite{1980ApJ...238..471R,1939LicOB..19...41B,Salucci:2018hqu}, galaxy clusters \cite{2011ARA&A..49..409A,1933AcHPh...6..110Z}, the early Universe \cite{Planck:2018vyg}, and gravitationally-lensed systems \cite{Clowe:2003tk} cannot be explained without the addition of new particles beyond the Standard Model. Despite a robust experimental program, dark matter has resisted attempts to measure its particle physics interactions, and astrophysical probes remain a vital window into its properties.

In this work, we employ a novel method to determine the dark matter density around the Solar location within the Milky Way, one that has never before been applied to data.
Our method is fully data-driven, uses the measurements of stellar position $\vec{x}$ and velocities $\vec{v}$ made possible by the \Gaia{} Space Telescope \cite{2021Gaia,2016A&A...595A...1G}, and is powered by modern, unsupervised machine learning methods.

The phase space density $f(\vec{x},\vec{v})$ of the population of stars within the Milky Way obeys the collisionless Boltzmann Equation:
\begin{equation}
\frac{\partial f}{\partial t}+v_i\frac{\partial f}{\partial x_i} = \frac{\partial\Phi}{\partial x_i}\frac{\partial f}{\partial v_i}  \label{eq:boltzmann_mastereq}.
\end{equation}
Here, $\Phi$ is the total gravitational potential, which can be related to the total mass density $\rho$ using the Poisson Equation
\begin{equation}
4\pi G \rho = \nabla^2 \Phi.
\label{eq:poisson_eq}
\end{equation}
Assuming that the phase space density of the stars is in equilibrium $\partial f/\partial t = 0$, the 3D acceleration field $-\partial \Phi/\partial x_i$ can be derived from knowledge of the stellar phase space density $f$ today. A further derivative of $\Phi$ then gives the total mass density $\rho$; the dark matter density can then be calculated assuming knowledge of the baryonic components. 

Measuring the stellar phase space density and its derivatives has traditionally been difficult, given the relatively high dimensionality (six) of the data. Instead, measurements of the local Galactic potential have used moments of the Boltzmann Equation -- the Jeans Equation -- along with simplifying assumptions (axisymmetry, specific functional forms, and/or small mixed radial and altitude ``tilt'' terms) which allows for relatively stable calculation of numeric derivatives from stellar data binned in the position coordinates. We refer to Refs.~\cite{2018MNRAS.478.1677S,2020A&A...643A..75S,2020MNRAS.494.6001N,2021ApJ...916..112N,2020MNRAS.495.4828G,2018A&A...615A..99H} for recent examples of these techniques.

The modern machine learning method known as normalizing flows provides a new approach to this problem that allows direct, unbinned access to the phase space density, independent of symmetry assumptions. Normalizing flows (reviewed in Ref.~\cite{9089305}) are a class of unsupervised deep learning algorithms that are sufficiently expressive to allow accurate modeling of the phase space density of high-dimensional data. Using normalizing flows, Refs.~\cite{2020arXiv201104673G,2021MNRAS.506.5721A,10.1093/mnras/stac153} directly solved the Boltzmann Equation for synthetic mock stellar data drawn from smooth analytic simulations of a galaxy \cite{2012ApJ...761...71Z,2012ApJ...758L..23L}. In Ref.~\cite{2023MNRAS.521.5100B}, we demonstrated this approach on a fully-cosmological $N$-body simulation of a Milky Way-like galaxy, including realistic \Gaia{}-like measurement errors and the impact of departures from symmetry and lack of equilibrium. 

Here, we apply the algorithm developed in Ref.~\cite{2023MNRAS.521.5100B} to \Gaia{} Data Release 3 (DR3) \cite{2022arXiv220800211G} itself. Using a population of stars within 4~kpc of the Sun for which full kinematic solutions are available, we measure the gravitational acceleration and total density everywhere within 3~kpc of the Solar location except in regions near the disk where dust extinction is significant. We also estimate the total uncertainty on our acceleration and density measurements throughout this region.
At each location, these uncertainties include statistical uncertainty, \Gaia{} measurement uncertainty, and an estimate of fit uncertainty from the normalizing flows. As the flow is extremely expressive, our errors should encompass a fuller range of possible shape variations of the density profile consistent with data, compared to many other approaches that fit the profile to a (perhaps overly-restrictive) functional form. Using existing measurements of the baryon density, we find clear evidence of a non-baryonic component to the mass density throughout the Solar neighborhood. 

Though these measurements do not rely on any assumptions of symmetry within the data, imposing spherical symmetry on the dark matter density allows us to average measurements at different locations and reduce errors. Under this additional assumption, we find a dark matter density of $0.47\pm0.05$~GeV/cm$^3$ at the Sun's distance from the Galactic Center. We also fit our density measurements to a generalized Navarro-Frenk-White (NFW) \cite{1996ApJ...462..563N,1997ApJ...490..493N} profile, though with considerable uncertainties on the best-fit parameters.

Future data releases from \Gaia{} will increase the number of stars with full kinematic information by a factor of $\sim 3$ \cite{2022arXiv220605902K}, as well as decreasing the proper motion measurement errors by $\sim 2$ \cite{gaia_science_performance_2022, 2022arXiv220800211G}. Combined with anticipated improvements in understanding the error model and quantifying dust extinction, the accuracy of the dark matter density measurements obtained using this method can be greatly increased in the near future.

In Section \ref{sec:data}, we introduce the \Gaia{} DR3 dataset used to train our normalizing flows. Section~\ref{sec:methods} contains the core results from our analysis: here we show our estimates of the phase space density using normalizing flows, followed by the calculations of accelerations and mass density using the collisionless Boltzmann Equation. Additionally, we investigate evidence for departures from equilibrium in the data and perform self-consistency checks.  In Section \ref{sec:discussion} we discuss our results and future directions for flow-based modeling of Galactic dynamics.

\section{Gaia DR3} \label{sec:data}

The \Gaia{} space telescope \cite{2016A&A...595A...1G,2022arXiv220800211G} has revolutionized precision astrometry. As of its third data release (DR3), \Gaia{} has measured the full 6D kinematics of nearly 33 million stars \cite{2022arXiv220605902K}.
This unprecedented volume of data, combined with state-of-the-art density estimation techniques, allows for robust mapping of the Milky Way's phase space density. 

For those stars with full kinematic information, the angular positions $(\alpha,\delta)$, proper motions on the sky $(\mu_\alpha^*,\mu_\delta)$, and parallax $\varpi$ are measured by the \Gaia{} photometer while the radial velocity $V_{\rm rad}$ and apparent magnitude $G_{\rm RVS}$ are measured by the \Gaia{} RVS spectrometer.
In regions with little dust extinction, nearly 100\% of stars with apparent magnitude brighter than $G_{\rm RVS} = 14$ are expected to have 6D kinematics in \Gaia{} DR3 \cite{2022arXiv220605902K}. In addition to a maximum apparent magnitude, the \Gaia{} RVS spectrometer also has a minimum apparent magnitude due to saturation \cite{2022arXiv220605725S}. Stars brighter than $G_{\textrm{RVS}} \approx 3$ are not included in the dataset used for this analysis.

In terms of $G_{\rm RVS}$ and the parallax-derived distance $(d/{\rm kpc})=(1~{\rm mas}/\varpi)$, the absolute magnitude $M_G$ is
 \begin{equation}
     M_G = G_{\rm RVS} - 5\log_{10}(d/{\rm kpc})-10.
 \end{equation}
In this analysis, we do not use other spectral information (such as $BP-RP$ color).

\begin{figure}[t!]
    \centering
    \includegraphics[width=0.6\columnwidth]{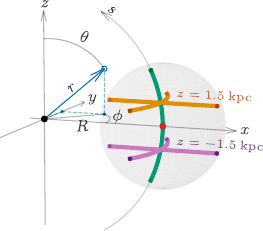}
    \caption{
        Schematic representation of the Solar location (red dot) relative to the Galactic Center (black dot). The 4~kpc observation volume is shown as a transparent grey sphere. The three coordinate systems used in this work are shown: the Galactocentric Cartesian coordinates $(x,y,z)$, the spherical coordinates $(r,\theta,\phi)$, and the cylindrical coordinates $(R,\phi,z)$. The lines through the observational volume with low dust extinction along which we measure accelerations and mass densities are shown in color. In orange, we show two lines at $z=+1.5$~kpc, one varying $r$ and another $\phi$. The two corresponding lines at $z=-1.5$~kpc are shown in purple. The line parameterized by polar arclength $s = r_\odot \times (\pi/2-\theta)$ passing through the Solar location is shown in green.
    }
    \label{fig:coordinate_systems}
\end{figure}

Given the approximate axisymmetry of the baryonic disk of the Galaxy and the approximate spherical symmetry of the dark matter distribution, it is useful to consider the data in both spherical and cylindrical coordinates, as well as Galactocentric Cartesian coordinates. The three coordinate systems we use are shown in Figure~\ref{fig:coordinate_systems} and are defined as follows:
\begin{enumerate}
\item Our Cartesian coordinate system places the $x-y$ plane in the Galactic disk, with $x=0$ at the Galactic center and the Sun along the $+x$ axis. The $+z$ axis (perpendicular to the disk) is oriented so that the net rotation of the disk stars in the $-y$ direction. 
\item The cylindrical coordinate system $(R,\phi,z)$ uses the same $z$ axis as the Galactocentric Cartesian coordinates. The Sun is located at $\phi=0$, with positive $\phi$ increasing towards the $+y$ axis. 
\item The spherical coordinate system $(r,\theta,\phi)$ has the same $\phi$ angle as the cylindrical coordinates, and measures $+\theta$ relative to $+z$ axis, with the Galactic disk at $\theta=\pi/2$. We define the polar arc length above or below the Galactic disk at the Solar radius (along the $\theta$ direction) as $s\equiv r_\odot\times(\pi/2-\theta)$, with $r_\odot = 8.122$~kpc.
\end{enumerate}

 From the fundamental kinematic properties (parallax, angular position, {\it etc.}) measured by \Gaia{}, the positions and velocities in these three coordinate systems can be obtained, using the parallax to calculate distance and assuming a Galactocentric Solar position and velocity of $(8.122,\,0.0,\,0.0208)$~kpc \cite{2018A&A...615L..15G,2019MNRAS.482.1417B} and $(-12.9,\,-245.6,\,7.78)$~km/s \cite{2018RNAAS...2..210D,2018A&A...615L..15G,2004ApJ...616..872R}, respectively. 

\begin{figure*}[t!]
    \centering
    \includegraphics[width=\textwidth]{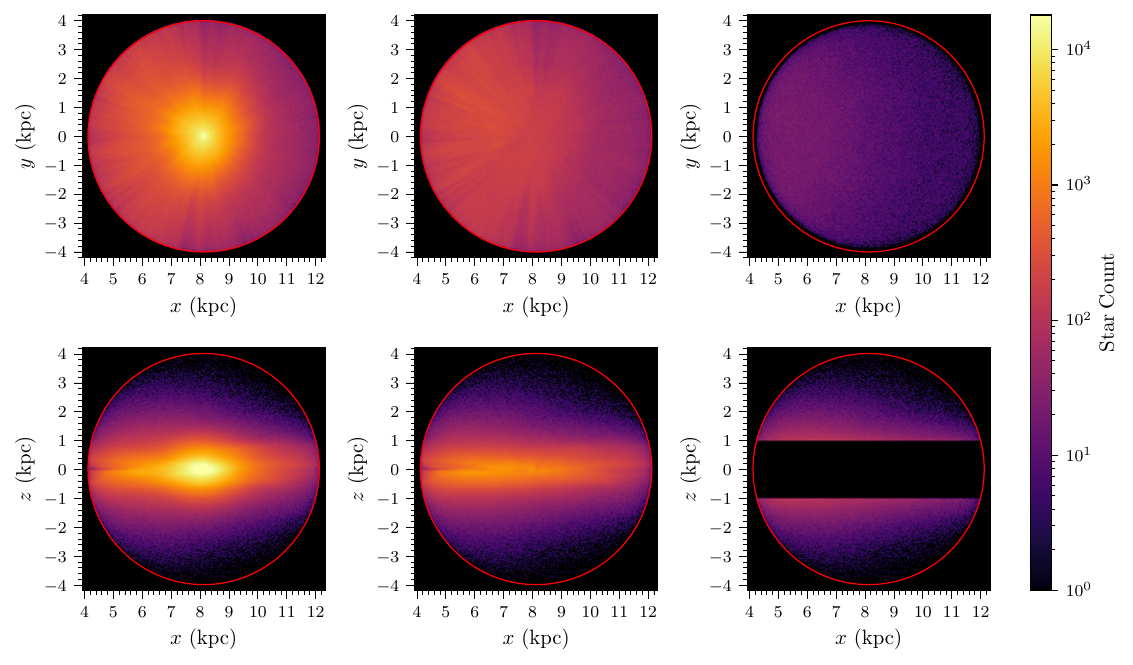}
    \caption{Density plots of the stars with full 6-dimensional kinematic information available from \Gaia{} within 4~kpc of the Solar location in the $x-y$ (top row) and $x-z$ (bottom row) planes. The left column shows all 24,789,061 fully-characterized stars. The middle column shows the 5,811,956 remaining stars after applying the selection criteria described in the text. The right column applies the additional requirement of $|z|>1$~kpc, resulting in 470,702 stars.
    }
    \label{fig:stars_numberdensity}
\end{figure*}

\begin{figure}[t!]
    \centering
    \includegraphics{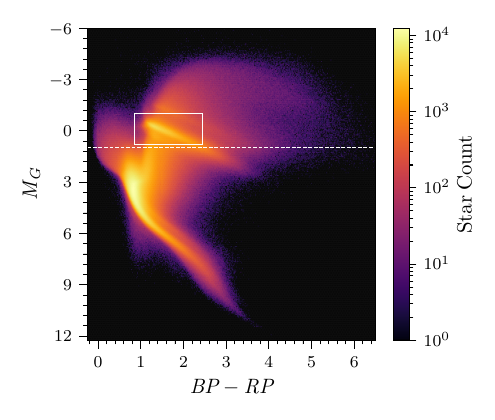}
    \caption{Absolute magnitude $M_G$ and color $BP-RP$ of all 24,789,061 stars within 4~kpc of the Sun with 6-dimensional kinematic information measured by \Gaia{}. The horizontal dashed white line denotes the magnitude completeness criteria Eq.~\eqref{eq:magcut}. All 5,811,956 stars above the dashed white line are bright enough to be observable for \Gaia{} regardless of position in the 4~kpc sphere. The visible peak in the white box is the red clump. All features in the space of uncorrected $M_G$ and $BP-RP$ will appear to be smeared towards the bottom-right of this figure, as dust extinction both dims and reddens stars. While a fraction of red clump stars are smeared beyond the magnitude cut threshold, all of these stars reside in the disk and will not significantly influence solutions to the CBE in the stellar halo.}
    \label{fig:cmd}
\end{figure}

\subsection{Tracer Population Selection}\label{sec:selection}

To extract the accelerations from the derivatives of stellar phase space density using the Boltzmann Equation, the population of stars in question must be complete, unbiased, and in dynamic equilibrium. These conditions are not satisfied by the full \Gaia{} 6D dataset, which is complete in observed (but not absolute) magnitude. Stars which are intrinsically dim but nearby are included, while intrinsically dim but more distant stars were not observed and thus are absent.

We first remove from our analysis dataset stars without spectroscopic or photometric magnitudes, as well as stars with large relative parallax errors ($3\sigma_\varpi > \varpi$).\footnote{This parallax error selection for the dataset quality control has the potential to bias the learned stellar number density, as further stars tend to have larger relative parallax errors. Nevertheless, the fraction of stars removed solely due to this selection criterion is at most 1.5\% within the fiducial volume of our analysis, and the statistical uncertainty of flow-learned number density is about 1\% around the Sun and about 4\% on the boundary of the fiducial volume; hence, the bias is insignificant. We discuss the effect of the relative parallax error selection in Appendix \ref{app:parallax_err_sel}. The explicit correction of this subleading bias in number density is beyond the scope of this paper, and we will leave it for future studies.} 
Of the 31,532,490 stars in  \Gaia{} DR3 with full 6D kinematic solutions available within 10~kpc, 29,855,114 stars remain after these cuts. We further remove 13 abnormally fast stars, with speeds higher than 1000~km/s relative to the Galactic rest frame. Within 4~kpc of the Sun, 24,789,061 stars remain, their position-space distributions are shown in the left column of Figure~\ref{fig:stars_numberdensity}, and their color-magnitude diagram ($BP-RP$ versus $M_G$) is shown in Figure~\ref{fig:cmd}.

In order to correct for the bias in the dataset towards stars with low observed magnitude, we require that every star in our restricted sample be bright enough to have been observed by \Gaia{} regardless of its position within a sphere around the Sun. We set this sphere to be 4~kpc in radius, both to limit fractional parallax errors (which grow with distance), as well as to allow a sufficiently large number of stars to pass the completeness criteria. That is, we require that the absolute magnitude $M_G$ of every star in our sample be bright enough that, if the star were located at 4~kpc, it would have an observed magnitude $G_{\rm RVS} < 14$ and thus be above the completeness limit of the \Gaia{} RVS spectrometer: 
\begin{equation}\label{eq:magcut}
    M_G < 14 - (5\log_{10}4+10).
\end{equation}
This selection criterion is shown as the white dashed line in Figure~\ref{fig:cmd}. Applying this selection removes 63.6\% of the stars in the sample with distance $<10$~kpc, leaving $10,876,430$. 
Note that we place this selection criterion on the magnitude without extinction-correction as the analysis of this paper will focus on the regions with less dust extinction.
This approach also helps mitigate issues arising from the non-uniform availability of dust extinction corrections from \Gaia{} across the 4~kpc sphere centered on the Sun. 
As a result of this non-uniformity, correcting for extinction induces additional spurious position-dependent suppression factors in the derived phase space density, depending on the availability of this information.
In the future, these corrections may be more uniformly available, and future versions of \Gaia{} itself and combined catalogs like \texttt{StarHorse} \cite{starhorse} would likely improve this analysis.

\begin{figure*}[t!]
    \centering
    \includegraphics[width=0.5\textwidth]{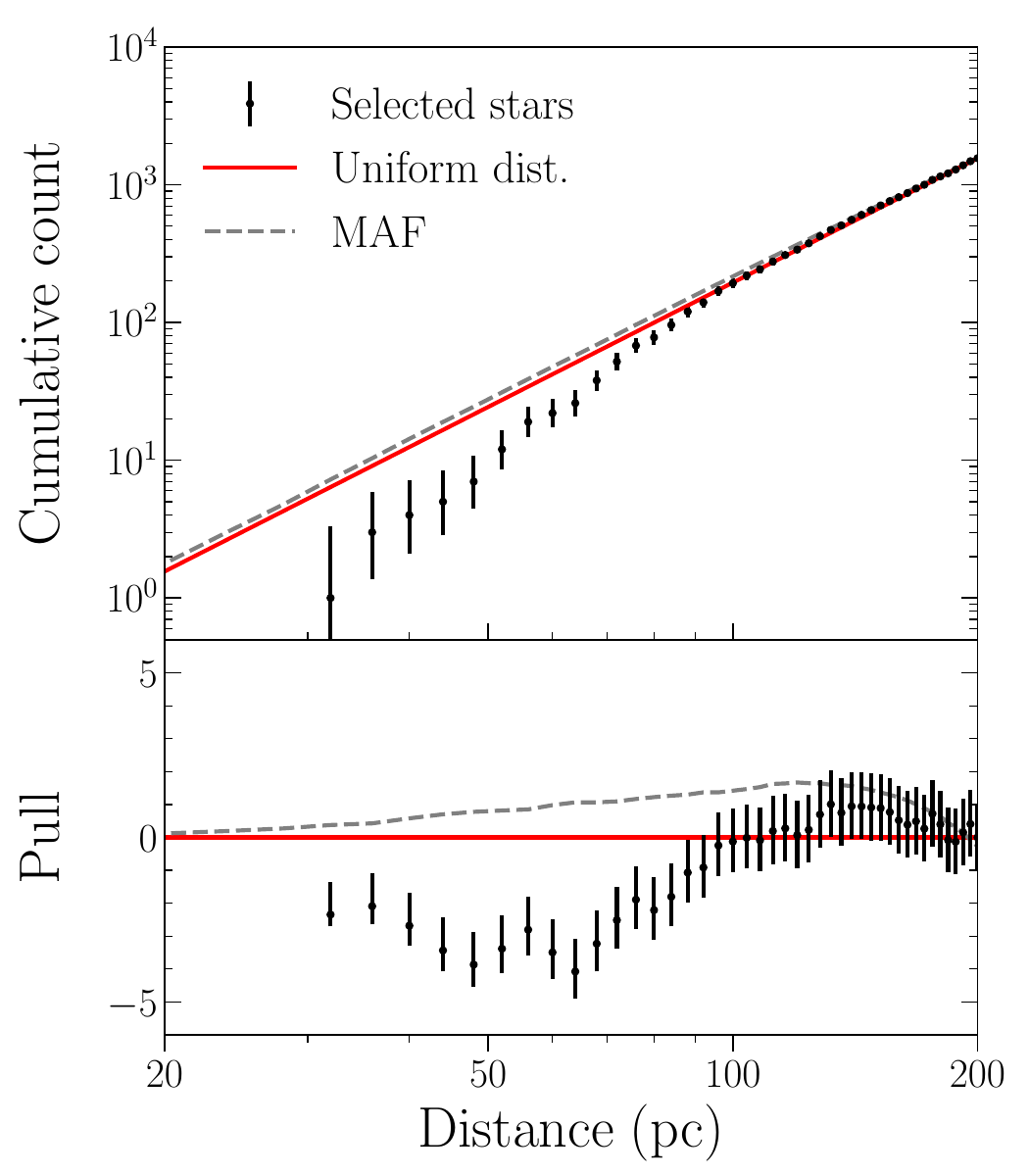}
    \caption{
        The cumulative histogram of distance to a star, $d$.
        Black points are the cumulative counts of selected stars given $d$, with error bars indicating $1\sigma$ statistical uncertainty.
        The red line represents the expected cumulative histogram for a uniform distribution, normalized to match the observed cumulative count at $d = 200$ pc.
        The gray dashed line shows the expected cumulative histogram of the flow-generated dataset, normalized to match the observed cumulative count at $d = 200$ pc.
        The lower panel shows the pull distributions, i.e., the difference between the uniform distribution histogram and the corresponding histogram divided by the $1\sigma$ statistical uncertainty of the dataset.
    }
    \label{fig:center_hole}
\end{figure*}

Due to the saturation limit $G_{\textrm{RVS}} \approx 3$, the closest star to the Sun in our final dataset is 30.4 pc away, and there are somewhat fewer stars within 50 pc in the analysis than would otherwise be expected.
Figure~\ref{fig:center_hole} shows the cumulative histogram of distance to a star, $d$, clearly revealing a small central hole where no stars are selected below $d < 30$ pc.

This scale of 30 pc is much smaller than the vertical length scale of the Milky Way's thin disk so we compare the histogram to a uniform distribution in order to understand the effect of the saturation limit in the hole's shape qualitatively.
The red line is the cumulative histogram of the uniform distribution with the same cumulative count at $d=200$ pc.
The deviation from uniform distribution begins below 80 pc.

We find that the normalizing flows, introduced in Section~\ref{sec:phasespace}, do not greatly suppress the phase space density of stars near the Solar location.
The cumulative histogram of flow-generated stars, shown as the gray dotted line in Figure~\ref{fig:center_hole}, more closely resembles the uniform distribution than the histogram of selected stars. 
This hole in the dataset is effectively smoothed out.
This smoothing is mainly due to the spectral bias of multilayer perceptrons \cite{pmlr-v97-rahaman19a}: multilayer perceptrons prioritize learning low-frequency features. 
While more expressive density estimation models \cite{NEURIPS2019_7ac71d43, song2021scorebased} and augmenting inputs with Fourier features \cite{NEURIPS2020_55053683} could enable complete learning of small-scale features, our focus is on estimating mass density averaged over a volume with length scale $\mathcal{O}[0.1]$ kpc using the same local mass density estimation setup developed using an $N$-body simulated galaxy \cite{2023MNRAS.521.5100B}.
We employ this selected stellar population with a central hole, and the flows will smooth it out implicitly.

For our final analysis, we require the measured distance of stars from the Sun to be less than 4~kpc, leaving 5,811,956 stars. The larger dataset (including stars out to 10~kpc) will be used when quantifying the impact of measurement errors on our final results, as we will discuss below.
In the middle column of Figure~\ref{fig:stars_numberdensity}, we show the stellar number density as a function of Galactocentric position after selecting on bright stars Eq.~\eqref{eq:magcut}.
The observational bias causing the stellar densities to be higher near the Sun is suppressed after the selection and the number density of stars is (correctly) seen to be rising towards the Galactic center. 

Despite this completeness cut on magnitude, incompleteness due to crowding and dust remains in select regions of the sky. In general, the effects of crowding are expected to be minimal for $G_{\textrm{RVS}} < 15$ (Ref.~\cite{2005A&A...430..129R}). However, our population of tracer stars is suppressed in the direction of Baade's Window, an extremely crowded field towards the Galactic center that is challenging for the $\Gaia{}$ RVS (Ref.~\cite{1999BaltA...8...97B}).

\begin{figure*}[ht]
        \includegraphics[height=1.25in, trim={0 0 0.7in 0}, clip]       {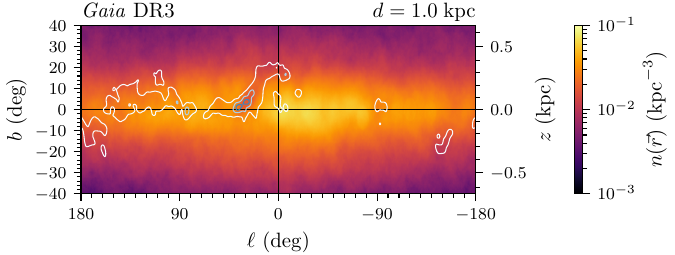}
        \includegraphics[height=1.25in]{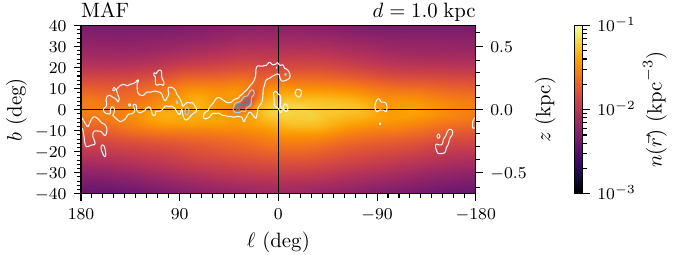}
        \includegraphics[height=1.25in, trim={0 0 0.7in 0}, clip]        {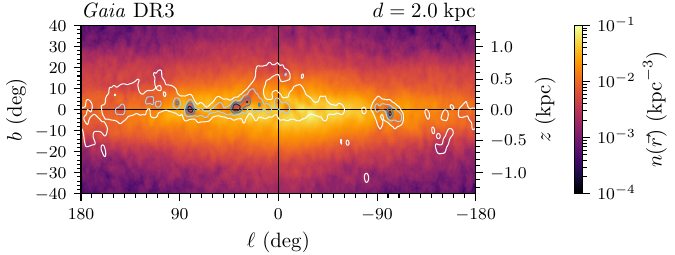}
        \includegraphics[height=1.25in]{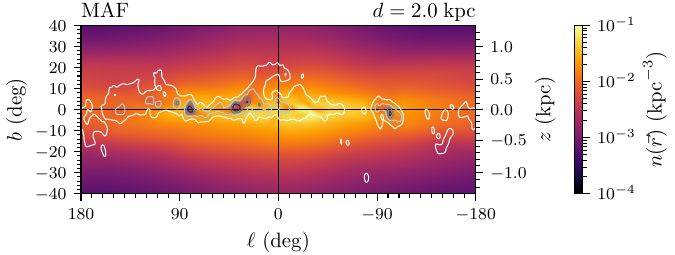}
        \includegraphics[height=1.25in, trim={0 0 0.7in 0}, clip]        {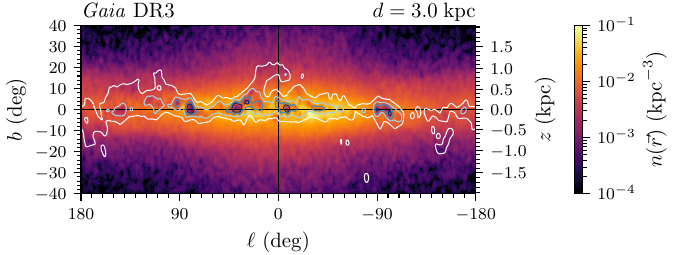}
        \includegraphics[height=1.25in]{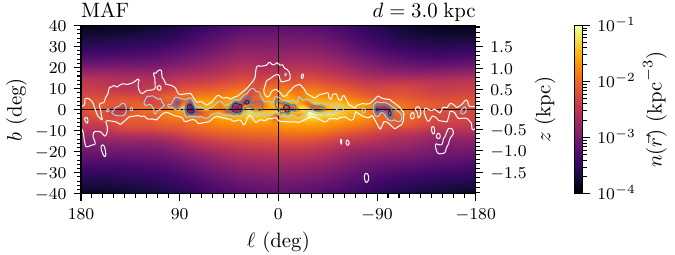}
    \caption{Stellar number density from \Gaia{} data (left) and MAF-learned number density (right) $n(\vec{r})$ on Galactic longitude ($\ell$) and latitude ($b$) planes at a distance (top) 1~kpc, (center) 2~kpc, and (bottom) 3~kpc.
    The stellar number densities are obtained by directly counting the number of stars within 0.1~kpc from the center of the pixel. The MAF-learned number densities are described in Section~\ref{sec:methods}.
    The $K_s$-band extinction maps at a given distance (obtained from the \texttt{dustmaps} package \cite{2018JOSS....3..695M}) are shown as contours. The maps are Gaussian-kernel smoothed with a bandwidth 
    of $8^\circ$. 
    We show four extinction value contours from white to blue: 0.15, 0.3, 0.45, and 0.60.   
    \label{fig:edge_on_dust}
    }
\end{figure*}

Within the dust-filled disk, significant density variations remain after the application of Eq.~\eqref{eq:magcut}. This is most obviously seen as a triangular wedge of apparent low stellar density towards the Galactic center at $|z|\sim 0$, but striations can also be seen in other directions within the disk. These features are primarily in the disk ($|z|\lesssim 1$~kpc) and largely trace known dust features, as can be seen in the left-hand column Figure~\ref{fig:edge_on_dust}, where we overlay extinction maps of dust \cite{2018JOSS....3..695M} with the observed number density of stars after the magnitude selection criterion has been applied. For this comparison, we stitch together two three-dimensional maps: \texttt{bayestar19} \cite{Green_2019} covering declination  $>30^{\circ}$, and \texttt{marshall} \cite{marshall} covering $b \in [-10^{\circ}, 10^{\circ}]$ and $\ell \in [-100^{\circ}, 100^{\circ}]$.
For the position not covered by either 3D dust maps, we use a 2D dust map \texttt{SFD} \cite{Schlegel_1998} for distances $d>1.0\; \mathrm{kpc}$. 
The regions of low stellar density match the known dust clouds, including the Serpens-Aquila rift, the Rho Ophiuchi cloud complex, Lupus, the Dark River, Pipe Nebula, the Northern Coalsack, and the Vela molecular ridge. In the right column of Figure~\ref{fig:stars_numberdensity}, we show the stars after the magnitude selection Eq.~\eqref{eq:magcut} with the additional requirement of $|z|>1$~kpc. Neither the effects of crowding nor dust are visible outside the disk. 

\begin{figure*}
    \begin{center}
        \includegraphics[width=0.45\columnwidth]{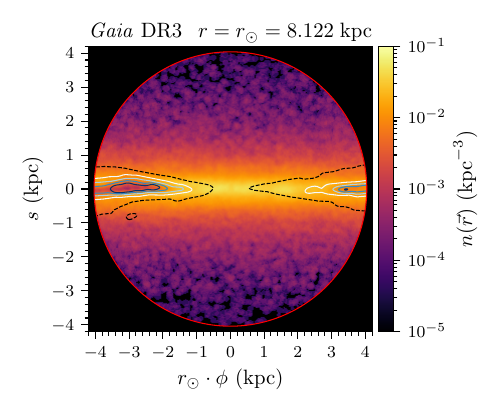}
        \includegraphics[width=0.45\columnwidth]{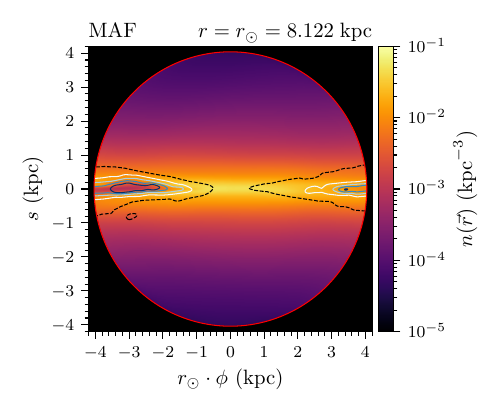}
    \end{center}
    \caption{
    Stellar number density from \Gaia{} data (left) and MAF-learned number density (right) as a function of $s$ and $r_\odot\cdot\phi$, at $r=r_\odot=8.122$~kpc. 
    The stellar number densities are obtained by directly counting the number of stars within 0.1~kpc from the center of the pixel. 
    The left-hand plot contains 206,852 stars.
    The MAF-learned number densities are described in Section~\ref{sec:methods}.
    We overlay the contour plot of three dimensional $K_s$ band extinction map obtained from the \texttt{dustmaps} package \cite{2018JOSS....3..695M}. The map is Gaussian-smoothed with a bandwidth 0.2~kpc.
    The dashed black contour is an extinction of 0.05, and the four contours from white to blue are extinction values of 0.15, 0.3, 0.45, and 0.60.
    }\label{fig:faceon_dust}
\end{figure*}

Although the effects of dust occlusion are greatest inside the disk ($|z|\lesssim 1$~kpc), one notable exception is in the neighborhood closest to the Sun: for the stars nearest to us, dust does not have enough opportunity to accumulate along the sight-lines, and so
extinction remains relatively low. As a result, measurements of phase space density along lines perpendicular to the disk and passing through the Solar location should be reliable and mostly unaffected by dust.  This is seen in Figure~\ref{fig:faceon_dust}, where we show the binned stellar densities as a function of $s$ and azimuthal angle $\phi$, along with contours of dust extinction.

The magnitude selection results in a sample of tracer stars that is unbiased by apparent magnitude within 4~kpc, and thus complete in absolute magnitude
(modulo the effects of dust and crowding). Roughly 69\% of these stars belong to the red clump, a sizable sub-population of the red giant branch (RGB) that is tightly clustered in the color-magnitude diagram (indicated by a white rectangle in Figure~\ref{fig:cmd}).
Red clump stars are typically between 1-4 Gyrs old \cite{2016ARA&A..54...95G}. Older stellar populations are preferred for kinematic studies of the Milky Way, as they have had sufficient time to equilibrate over the Galaxy's dynamic timescales \cite{2014JPhG...41f3101R}. Stars from the red clump meet this criteria, and have been used in a recent precision Jeans analysis of the Solar neighborhood \cite{2020A&A...643A..75S}.

In principle, the percentage of the sample composed of red clump stars could be increased by selecting based on extinction-corrected color in addition to $M_G$. However, similar to extinctions for the \Gaia{} RVS spectrometer, extinction-corrected colors are not uniformly available across the sky. Using corrected colors would lead to selection effects in the phase space density and errors in the solution to the Boltzmann equation. While we expect this to be less of an issue in future data releases from \Gaia{}, for this analysis we do not apply color selection criteria.

We will apply our normalizing flow algorithm to the complete, unbiased dataset within 4~kpc, but with the knowledge that gravitational accelerations or mass densities within the disk far from the Sun are not reliable due to dust extinction. In Figure~\ref{fig:coordinate_systems}, we show the dust-avoiding lines along which we measure accelerations and mass densities.  

\subsection{Measurement Errors} \label{subsec:measerror}

\begin{figure*}[ht!]
    \includegraphics[width=\textwidth]{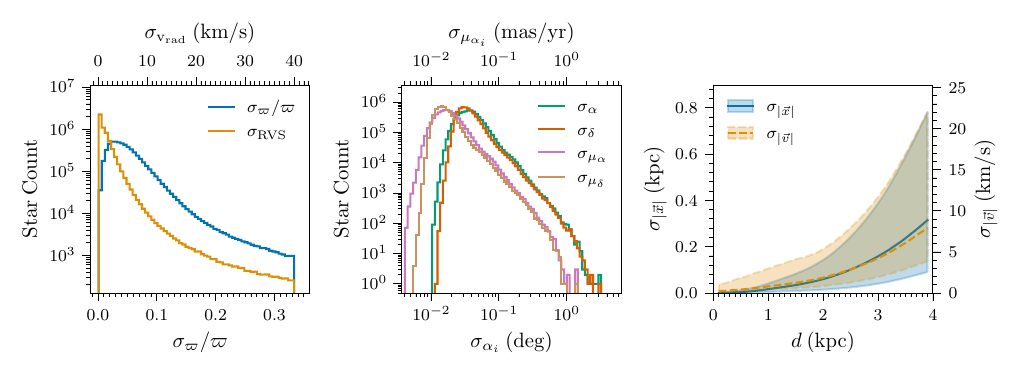}
    \caption{Left and Center: $1\sigma$ standard deviations (as reported by \Gaia{} DR3) of the measured kinematic parameters for stars passing the distance, magnitude, and parallax error selections of Section~\ref{sec:selection}. Errors are shown for parallax $\varpi$ (left plot, lower axis), radial velocity $v_{\rm rad}$ (left plot, upper axis), angular ICRS position $(\alpha,\delta)$ (center plot, lower axis), and ICRS proper motion $(\mu_\alpha, \mu_\delta)$ (center plot, upper axis). We discard stars with relative parallax errors larger than 1/3. Stars with radial velocity uncertainties larger than 40~km/s are not included in \Gaia{} DR3 \cite{2022arXiv220605902K}. Right: Median standard deviation in Cartesian position (blue line, left vertical axis) and velocity (yellow line, right vertical axis) as a function of stellar distance $d$. The $16^{\rm th}$ and $84^{\rm th}$ percentiles of these standard deviations are shown as asymmetric bands around the central median value.
    } 
    \label{fig:errors_measured}
\end{figure*}

\Gaia{} DR3 provides measurement errors in the form of Gaussian standard deviations for the measured quantities of angular position, parallax, proper motion, RVS, and $G_{\rm RVS}$. In the left and center panels of Figure~\ref{fig:errors_measured}, we show histograms of the kinematic errors for all stars within 4~kpc which pass the selection criteria described in Section~\ref{sec:selection}. Propagating the errors from the angular positions, proper motions, parallax, and RVS measurements results in a covariance matrix for the measurement errors in the Cartesian coordinates used for training the normalizing flows. The median standard deviation in the distance and speed of stars perturbed by the Cartesian error model is shown in the right panel of Figure~\ref{fig:errors_measured}.

The parallax and RVS measurements contribute the dominant sources of error to the kinematic solutions. In particular, parallax errors in some cases can be larger than the measured parallax; this often results in negative parallaxes when varying within errors. Considerable literature exists on the conversion of parallax to distance, including the impact of measurement errors \cite{2018A&A...616A...9L,2018AJ....156..145A,2016ApJ...832..137A,2019MNRAS.489.2079L,2019MNRAS.487.3568S}. We note that \Gaia{} also provides a secondary distance estimate 
\textsc{distance\_gspphot}, which is inferred from 
a Markov Chain Monte Carlo algorithm that uses spectra from the BP and RP bands, apparent $G$ magnitude, and parallax (\cite{2022arXiv220606138A,2013A&A...559A..74B}). Unfortunately, the availability of \textsc{distance\_gspphot} (as well as other data products from this fit) is not uniform across the sky. Given this limitation, we find the most uniform reliable distance estimate to be the inverse parallax after stars with large relative parallax errors are removed.

Our error propagation procedure follows the outline of Ref.~\cite{2023MNRAS.521.5100B}: we generate multiple variations of the original dataset after varying every star's kinematic features within their Gaussian errors, using the correlation matrix given the errors provided by \Gaia{} DR3 converted into Cartesian coordinates. For each varied iteration of the data, we calculate first the phase space density, then the gravitational acceleration and mass density (using the algorithms described in Section~\ref{sec:methods}). The variations of these quantities over multiple error-smeared datasets provide an estimate of the impact of measurement errors on the derived quantities. In order to avoid the edge effect of stars migrating out of the 4~kpc-radius sphere when errors are applied without a corresponding inward migration of more distant stars, for each variation of the dataset we apply the error-smearing to stars in the 10~kpc-radius sphere. After applying the errors to the larger dataset, we select those stars whose error-smeared distances place them within 4~kpc of the Sun.

\section{Accelerations and Mass Densities from Normalizing Flows}\label{sec:methods}

The techniques used in this work to calculate accelerations and mass density from the CBE using a flow-derived phase space density are described fully in Ref.~\cite{2023MNRAS.521.5100B}. We describe each component of our analysis briefly here, and present results using \Gaia{} DR3.

\subsection{Phase Space Densities}\label{sec:phasespace}

\begin{figure*}[ht!]
    \begin{center}
        \includegraphics[width=0.31\textwidth]{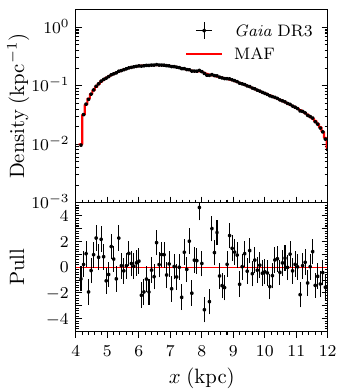}
        \includegraphics[width=0.31\textwidth]{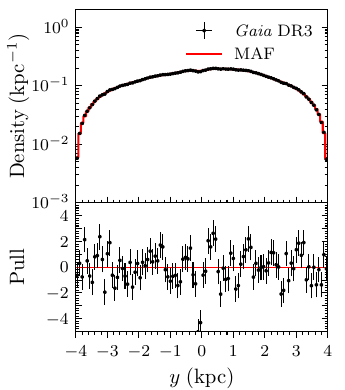}
        \includegraphics[width=0.31\textwidth]{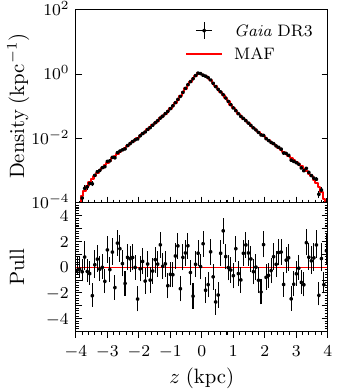}
    \end{center}
    \begin{center}
        \includegraphics[width=0.31\textwidth]{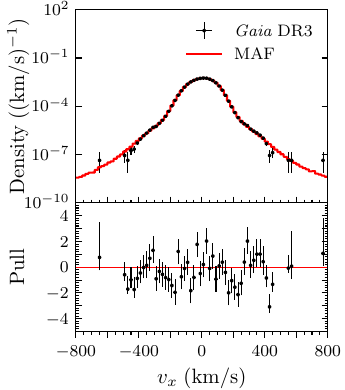}
        \includegraphics[width=0.31\textwidth]{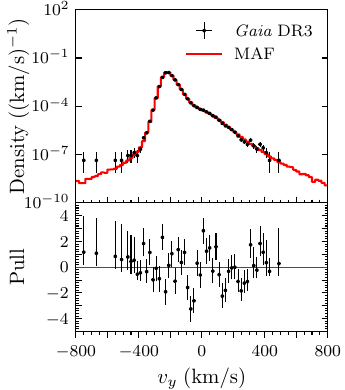}
        \includegraphics[width=0.31\textwidth]{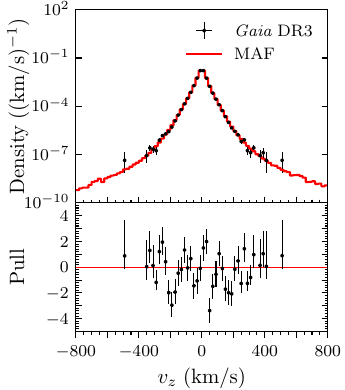}
    \end{center}
    \caption{
    Normalized histograms of (top) position components and (bottom) velocity components for selected stars in \Gaia{} DR3 (downsampled to 20\% of the original size).
    The red lines are the histograms for synthetic stars sampled from the normalizing flows.
    The error bars are the $1\sigma$ statistical uncertainty.
    Below the main plots, we show the pull distributions, i.e., the difference between \Gaia{} and MAF histograms divided by the $1\sigma$ statistical uncertainty. 
    }
    \label{fig:flow_density}
\end{figure*}

To determine the gravitational accelerations $-\vec\nabla \Phi$ from the collisionless Boltzmann Equation applied to a stellar population, we must first determine the phase space density $f(\vec{x},\vec{v})$ for that population. 
We accomplish this using normalizing flows, an unsupervised machine-learning technique for density estimation.
Normalizing flows are based on invertible transformations of a simple base distribution (such as standard normal distribution) into a more complicated distribution.
As long as the transformation is expressive enough, normalizing flows are able to model a variety of distributions so that this model can be used as a free-form density estimator. 
This expressivity is generally achieved using neural networks with bijective constraints.

We use two normalizing flows to model the stellar number density $n(\vec{x})$ and the conditional velocity distribution $p(\vec{v} | \vec{x})$ separately:
\begin{equation}
    f(\vec{x},\vec{v}) = n(\vec{x}) p(\vec{v} | \vec{x}).
\end{equation}
Note that the estimation of $p(\vec{v} | \vec{x})$ requires conditional density estimation; this is easily implemented in the normalizing flows architecture by simply making the transformation conditioned on the position vector $\vec{x}$.
In addition to increasing the accuracy of the overall phase space density, this decomposition is also helpful for solving the Boltzmann equation. 
In both cases, the loss functions $\mathcal{L}_{x}$ and $\mathcal{L}_{v}$ for training the density models $n(\vec{x})$ and $p(\vec{v} | \vec{x})$ are the negative log-likelihoods of the data: 
\begin{eqnarray}
    {\cal L}_{x} & = & - \frac{1}{N} \sum_{i=1}^N \log n(\vec{x}_i) \\
    {\cal L}_{v} & = & - \frac{1}{N} \sum_{i=1}^N \log p(\vec{v}_i|\vec{x}_i),
\end{eqnarray}
where $N$ is the size of an input dataset.

In this work, we implement the normalizing flows using the Masked Autoregressive Flow (MAF) \cite{NIPS2017_6c1da886} architecture as the transformation model. 
GELU activation \cite{2016arXiv160608415H} is used in order to model a smooth differentiable transformation.
We use the ADAM optimizer \cite{2014arXiv1412.6980K} to train the flows. 
We construct a validation dataset using 20\% of the stars, randomly selected.
Early stopping with a patience of 50 epochs is used, and we select the model with the lowest validation loss.
For the central values, we use Monte Carlo cross-validation and ensemble averaging in order to fully utilize the dataset and reduce noise in our density estimation.
That is, we prepare 100 different random splits of training and validation datasets and train a MAF for each.
The density is estimated by ensemble averaging the probabilities given by each of the MAFs. 
All the neural networks are implemented using \textsc{PyTorch} \cite{NEURIPS2019_bdbca288} and \textsc{nflows} \cite{nflows}. 
The details of our neural network architectures and data preprocessing prior to training are identical to those described in Ref.~\cite{2023MNRAS.521.5100B}.

In Figure~\ref{fig:flow_density}, we plot one-dimensional histograms of the selected data compared to the density estimated by the normalizing flows. Black markers are the histogram of selected stars randomly downsampled to 20\% of the original size (this matches the size of the randomly-selected validation dataset). 
Red lines are the histograms of synthetic stars generated by sampling from the MAFs. For this figure, we upsample from the MAF to 100 times the size of the \Gaia{} dataset.

The small deviations near the Solar location in $x$ and $y$ histograms are due to dust extinction in the disk (as seen in Figures~\ref{fig:stars_numberdensity} and \ref{fig:edge_on_dust}). 
Along lines of sight with significant dust, there are sharp falloffs in the apparent stellar density which do not reflect the true density of stars. 
The MAFs are constrained to be smooth functions, and as a result, they have difficulties modeling such discontinuities, and the quality of density estimation may degrade. 
The bump in the $x$ histogram is mainly due to dust clouds in 
Cygnus ($\ell \sim 75^{\circ}$) and Vela ($\ell \sim -95^{\circ}$),
while the bump in the $y$ histogram is due to dust clouds closer to the Galactic center.

The right-hand column of Figure~\ref{fig:edge_on_dust} shows the estimated number density $n(\vec{r})$ on the Galactic longitude and latitude $(\ell,b)$ planes at different distances $d$ from the Sun, again with the dust map overlaid.
Comparing to the observed binned number densities (left column of Figure~\ref{fig:edge_on_dust}), we can see that the MAF is correctly learning the overall density scales and substructures visible in the stellar counts within the disk, though detailed comparison suggests that the MAF may have difficulty replicating the small-scale sharp features of the dust.
MAFs are known to generate spurious ``wrinkles'' around sharp edges or topologically non-trivial structures within data \cite{huang2018neural}.
Though the wrinkles in density themselves are small, numerical artifacts will be amplified in density derivative estimations, resulting in biases in the acceleration and mass density estimations.
As previously mentioned, we avoid low-$|z|$ regions which are far from the Solar location to minimize the effects of dust on our measurements of the acceleration and mass density fields.
Density estimators that improve over the MAF results are available \cite{2022arXiv221111765L, song2021scorebased, lipman2023flow, tong2024improving}, but are computationally expensive for the density derivative estimations; we leave studies of such architectures for future work.

\subsection{Accelerations}\label{sec:accel}

Using the learned phase space density, we estimate the acceleration field $\vec a = -\vec{\nabla}\Phi$ by solving the collisionless Boltzmann equation. As observed in Refs.~\cite{2021MNRAS.506.5721A,2023MNRAS.521.5100B}, since $\Phi$ is a function of position only, we can approximately solve the Boltzmann Equation for the acceleration by the least squares method using generated velocity samples at a given $\vec{x}$ and minimizing the residual $\partial f/\partial t$. That is, at a given position $\vec{x}$, we calculate the accelerations from the trained MAFs by finding the value of $\vec{a}(x)$ that minimizes the following mean square error (MSE):
\begin{equation}
\mathcal{L}_a = \int d^3 \vec{v} \, p(\vec{v}|\vec{x}) \, 
\left| 
    v_i\frac{\partial f}{\partial x_i} +a_i(\vec{x})\frac{\partial f}{\partial v_i} 
\right|^2. \label{eq:mse}
\end{equation}
We evaluate this integral by quasi-Monte Carlo integration \cite{2023MNRAS.521.5100B} and minimize it analytically to determine the best-fit acceleration value. For each position, we sample 10,000 velocities with $|\vec{v}| < 600\;\mathrm{km/s}$ in order to obtain a stable acceleration solution with small statistical errors.

Given that Eq.~\eqref{eq:mse} is highly overconstrained, the residual $\partial f/\partial t$ is not guaranteed to be zero. However, our minimization procedure follows (in spirit) the approximation typically made in density measurements based on the Boltzmann Equation, which assume the phase space density $f$ is in equilibrium ($\partial f/\partial t=0$). Within the Milky Way this assumption would imply an axial and $z$-symmetric potential (and thus mass density and acceleration fields which respect these symmetries). There is evidence that the Milky Way is not in dynamic equilibrium \cite{2012ApJ...750L..41W,2013MNRAS.436..101W,2018Natur.561..360A}. As we do not enforce these symmetries in our acceleration calculation and since our MSE minimization can result in residual non-zero $\partial f/\partial t$, we can perform closure tests to estimate the amount of deviation from equilibrium and validate whether the accelerations are derived in a self-consistent manner. We discuss this in detail in Section~\ref{sec:equilibrium}.

\begin{table}[t]
\begin{center}
\begin{tabular}{rcc}
    \toprule
    &
    \Gaia{} EDR3 \cite{gaia_edr3_accel}&
    \textbf{This work}
    \\
    \midrule
    $a_x$ $(10^{-10}\mathrm{m/s^2})$ &
    $-2.32 \pm 0.16$ & 
    $\aX \pm \aXerr$ 
    \\
    $a_{y}$ $(10^{-10}\mathrm{m/s^2})$ & 
    $\phantom{-}0.04 \pm 0.16$ & 
    $\phantom{-}\aY \pm \aYerr$  
    \\
    $a_z$ $(10^{-10}\mathrm{m/s^2})$ & 
    $-0.14 \pm 0.19$ & 
    $\aZ \pm \aZerr$  
    \\
    $|\vec{a}|$ $(10^{-10}\mathrm{m/s^2})$ &
    $\phantom{-}2.32 \pm 0.16$ &
    $\phantom{-}\aMag \pm \aMagerr$
    \\
    \bottomrule
\end{tabular}
\end{center}
\caption{Galactic acceleration at the Solar location $\vec{a}_\odot$ in Cartesian coordinates, calculated by averaging the solution to the Boltzmann equation within a 100~pc sphere centered on the Sun. We list for comparison the acceleration at the Solar location obtained from \Gaia{} DR3 quasar measurements \cite{gaia_edr3_accel}.  \label{tab:acceleration:sol}}
\end{table}

\begin{figure*}[ht!]
    \begin{center}
        \includegraphics[width=0.32\textwidth]{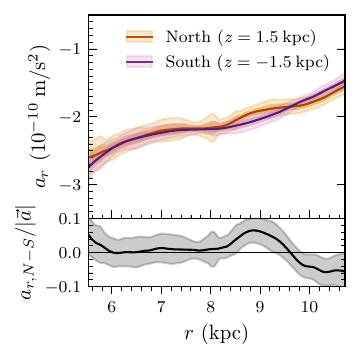}
        \includegraphics[width=0.32\textwidth]{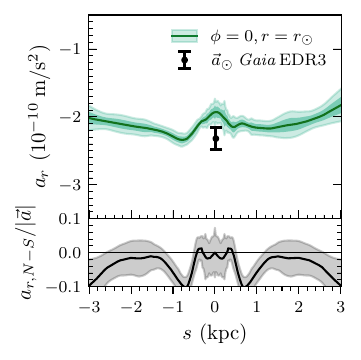}
        \includegraphics[width=0.32\textwidth]{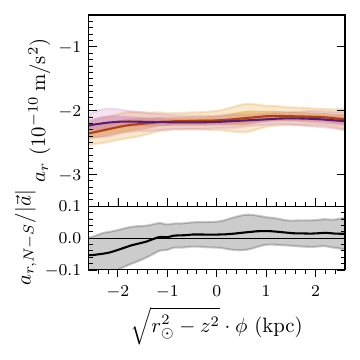}
    \end{center}
    \begin{center}
        \includegraphics[width=0.32\textwidth]{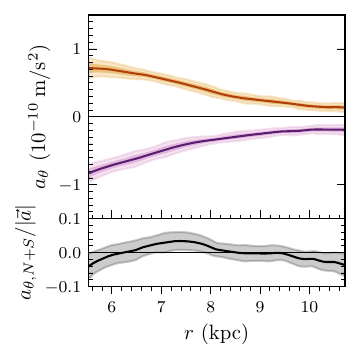}
        \includegraphics[width=0.32\textwidth]{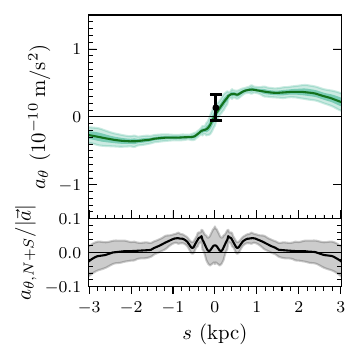}
        \includegraphics[width=0.32\textwidth]{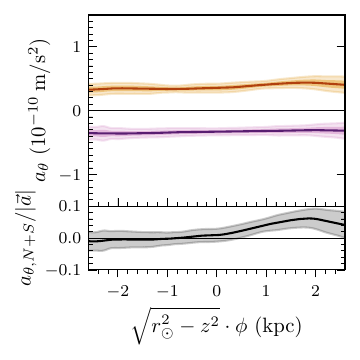}
    \end{center}
    \begin{center}
        \includegraphics[width=0.32\textwidth]{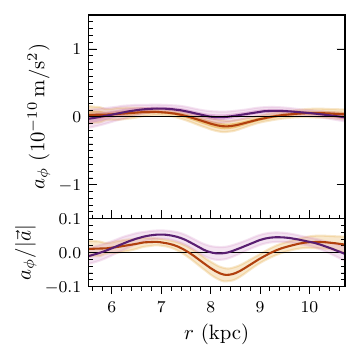}
        \includegraphics[width=0.32\textwidth]{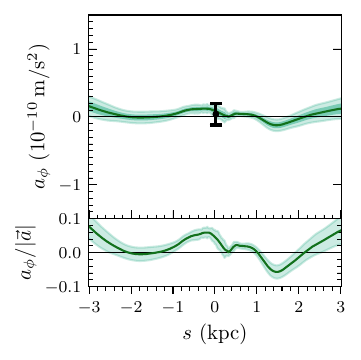}
        \includegraphics[width=0.32\textwidth]{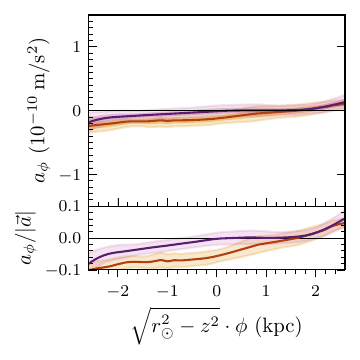}
    \end{center}
    \caption{Accelerations $a_r$ (top row), $a_\theta$ (middle row), and $a_\phi$ (bottom row) as a function of spherical radius $r$ (left column), polar arclength $s=r_\odot\times(\pi/2-\theta)$ (middle column), and azimuthal arclength $(r_\odot^2-z^2)^{1/2}\times\phi$ (right column). Radial and azimuthal measurements are taken along lines off-set from the Galactic midplane, at $z=1.5$~kpc (yellow lines, labeled ``North'') and $z=-1.5$~kpc (purple lines, labeled ``South''). See Figure~\ref{fig:coordinate_systems} for the orientation of these measurement lines within the Galaxy. The inner dark bands and outer light bands denote $1\sigma$ and $2\sigma$ uncertainties respectively. Subpanels in the top and middle rows show the fractional difference between the North and South measurements (appropriately mirrored for $a_\theta$). Subpanels in the bottom row show the ratio of $a_\phi$ to the magnitude of $\vec{a}$.
    }
    \label{fig:accelerations}
\end{figure*}

In Table~\ref{tab:acceleration:sol}, we show the estimated acceleration at the Solar location, averaged over a ball centered at the Sun and with a radius 100~pc.\footnote{Recall that our data has a small void at the Solar location with radius $\sim 50$~pc, due to the \Gaia{} lower limit on the apparent magnitude. Although our normalizing flows have smoothed out this region, we apply this averaging in order to suppress any artifacts from the interpolation.}
The measured radial acceleration at the Solar location is $(\aR \pm \aRerr) \times 10^{-10}\;\mathrm{m/s^2}$; the other components are negligible in comparison.
The error includes measurement and statistical errors estimated by resampling and bootstrapping, the variation from multiple independent trainings of the MAFs, and the standard deviation of the estimated acceleration within the 100~pc ball.

Our result agrees with the acceleration measurement of \Gaia{} EDR3  \cite{gaia_edr3_accel} within $2\sigma$. 
This latter measurement is obtained from the proper motion of quasars caused by secular aberration due to the orbital motion of the Solar system in the Milky Way. 
Differences between these two measurement techniques may indicate local disequilibrium in the Solar neighborhood, as we discuss in Section~\ref{sec:equilibrium}. 
In the future, the dataset available to \Gaia{} DR4 or DR5 may be sufficient to resolve any differences at a statistically significant level.

In Figure~\ref{fig:accelerations}, we show 
the estimated accelerations in spherical coordinates ($a_r$, $a_\theta$, $a_\phi$), measured along curves of $r$, $\phi$, and $s=r_\odot\cdot(\pi/2-\theta)$ through the volume around the Sun (avoiding regions of significant dust where the MAF-based acceleration estimate may be inaccurate).  The orientations of these measurement curves within the Galaxy are summarized in Figure~\ref{fig:coordinate_systems}, and we describe them further here:
\begin{itemize}
\item Accelerations as a function of $r$ are measured along lines passing through $(x,y,z)=(8.122,0,\pm 1.5)\,$kpc -- that is, through points 1.5~kpc above (towards Galactic North) and below (towards Galactic South) the Galactic disk at the Solar location. 

\item Accelerations as a function of $\phi$ are calculated along curves located 1.5~kpc North and South of the Solar location passing through $(x,y,z)=(7.982,0,\pm 1.5)$, constrained so that $r=r_\odot$ along the curve. These 1.5~kpc offsets above and below the disk are chosen to avoid the dominant sources of dust extinction and resulting wrinkles in the MAF number densities, as seen in Figures~\ref{fig:stars_numberdensity} and \ref{fig:edge_on_dust}. 

\item We measure accelerations as a function of the polar arc $s$ passing through the Solar location. These are only acceleration measurements we make closer to the disk than $|z|=1.5$~kpc. Sight-lines to this arc pass through the disk only near to the Earth, where extinction is less of a concern (see Figure~\ref{fig:faceon_dust}). For accelerations within 150~pc of the Sun, we average over a sphere (as in Table~\ref{tab:acceleration:sol}) to smooth over dust within the disk. No averaging is applied to points further away.
\end{itemize}

\subsection{Testing Equilibrium} \label{sec:equilibrium}

Our method's
flexibility and
lack of imposed symmetry allows us to perform self-consistency checks that are not usually possible with previous approaches for measuring acceleration and mass density based on the Boltzmann Equation or its moments. We consider two such checks here.

First, we note that the acceleration field calculated using the MAF minimizes $\partial \ln f/\partial t$ at each point in space. The MAF-derived acceleration $\vec{a}(\vec{x})$ represents the solution most consistent with equilibrium based on samples from $p(\vec{v}|\vec{x})$. In a perfectly equilibrated system, $\partial f/\partial t \equiv 0$ for all $\vec{v}$ where $p(\vec{v}|\vec{x})>0$. However, since the Milky Way is not in perfect equilibrium at any given location $\vec{x}$, the three degrees of freedom in $\vec{a}(\vec{x})$ are unlikely to find a perfect solution for the thousands of sampled velocities from $p(\vec{v}|\vec{x})$. The residual $\partial \ln f / \partial t$ from this fit allows us to quantify how much the equilibrium assumption deviates for the local Milky Way at any given $\vec{x}$.

\begin{figure}
    \centering
    \includegraphics{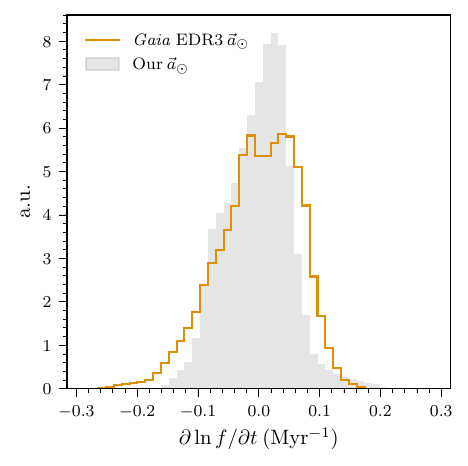}
    \caption{
    Distribution of $\partial \ln f / \partial t$ at the Solar location using the MAF-derived Solar acceleration $a_\odot$ (blue) and the \Gaia{} EDR3 quasar-derived measurement of $a_\odot$ (orange).
    }
    \label{fig:solar_boltzmann_closure}
\end{figure}

We test this at the Solar location by sampling $10^6$ velocities from the MAF model of $p(\vec{v}|\vec{x}_\odot)$ and computing the corresponding phase-space gradients. We calculate the acceleration $\vec{a}_\odot$ most consistent with equilibrium (identical to the Solar acceleration given in Table~\ref{tab:acceleration:sol}), and examine the residual $\partial \ln f/\partial t$. The resulting distribution of $\partial \ln f/\partial t$, shown in Figure~\ref{fig:solar_boltzmann_closure}, peaks at zero as expected, with a full width at half maximum (FWHM) of $0.13\:{\rm Myr^{-1}}$. The width of this distribution, given the units of $\partial \ln f/\partial t$, can be interpreted as a dynamical time $t_{\rm dyn} =1/ \sqrt{|(\partial \ln f/\partial t)^2|_{v}}$, which is large when $\partial \ln f/\partial t$ is small. For the MAF-derived $\vec{a}_\odot$, we find $t_{\rm dyn} = 8.0\:{\rm Myr}$. Since this estimate of $\vec{a}_\odot$ minimizes $\partial \ln f/\partial t$ by construction, this value represents the maximum possible $t_{\rm dyn}$ for any solution of $\vec{a}_\odot$.

To interpret the significance of $t_{\rm dyn}$, we compare our MAF-derived $\vec{a}_\odot$ to previous estimates of the Solar acceleration. Ref.~\cite{gaia_edr3_accel} used the proper motions of distant extragalactic quasars from \Gaia{} EDR3 to estimate $\vec{a}_\odot$. By replacing our MAF solution for $\vec{a}_\odot$ with the EDR3 value in the CBE while keeping the same phase-space gradients, we can assess the consistency of the EDR3 solution with the equilibrium assumption. Interestingly, as shown in Figure~\ref{fig:solar_boltzmann_closure}, this independent estimate of $\vec{a}_\odot$ is also consistent with $\partial \ln f/\partial t = 0$. The central value of $\partial \ln f/\partial t$ does not shift significantly despite the $\sim 20 \%$ difference between the two estimates of $\vec{a}_\odot$. However, we observe a clear shift in the widths of each distribution: the FWHM for the EDR3 solution is $0.16\:{\rm Myr^{-1}}$. This corresponds to a dynamical time of $t_{\rm dyn} = 6.4\:{\rm Myr}$, $20\%$ shorter than $t_{\rm dyn}$ from the MAF-derived $\vec{a}_\odot$.

Second, we note that departures from axisymmetry in the rotating Milky Way, would presumably imply disequilibrium. In the absence of measurement errors or disequilibrium and assuming maximally expressive MAFs, the northern ($z>0$) and southern ($z<0$) measurements of $a_r$ should be equal, and the $a_\theta$ values should be equal up to a negative sign. If the system respects axisymmetry, $a_\phi$ should be zero. In the subpanels of Figure~\ref{fig:accelerations}, we show the deviations from North-South symmetry for $a_r$ and $a_\theta$ over the magnitude of the acceleration, as well as $a_\phi/|\vec{a}|$. As can be seen, these measures of disequilibrium and/or departure from axisymmetry are at most $10\%$, suggesting that the system is in equilibrium at least to this level.

To summarize, under these closure tests, we find that our assumption of equilibrium in the analysis of the Boltzmann Equation is good to within the $\sim 10\%$ level, as quantified by deviations from axisymmetry and north-south reflection symmetry in the measured accelerations. Using an independently determined acceleration at the Solar location \cite{gaia_edr3_accel}, we also find closure (self-consistency) of our equilibrium assumption to within the level of precision of the MAF density estimation.  
In the future, larger datasets, better control of errors, and more expressive normalizing flow architectures may allow tighter distributions of $\partial\ln f/\partial t$, allowing statistically significant measures of equilibrium and axisymmetry.

\subsection{Mass Densities} \label{sec:density}

Given an acceleration calculated at every point in the volume around the Sun, the total mass density $\rho$ can be calculated using the Poisson equation in Eq.~\eqref{eq:poisson_eq}. This requires taking an additional numeric derivative of $\Phi$. Again following the algorithm of Ref.~\cite{2023MNRAS.521.5100B}, we calculate the second derivative of $\Phi$ at position $\vec{x}$ by convolving the accelerations over a truncated Gaussian kernel $K$ centered at $\vec{x}$:
\begin{equation}
4 \pi G\, \rho * K = (\nabla^2 \Phi) * K.
\end{equation}
Here $*$ indicates convolution with the kernel. The Gaussian kernel is truncated at $|\vec{x}/\vec{h}| =2$ where $\vec{h} = (0.5,0.5,0.2)$~kpc. This ellipsoidal kernel averages the mass density at scales below $\vec{h}$, and thus we are not sensitive to density fluctuations at  scales smaller than this. We draw 3,200 points to calculate the mass density at each $\vec{x}$, again using quasi-Monte Carlo sampling.

With the total mass density calculated at a point, we can then extract the dark matter mass density using a model for the baryonic components of the Milky Way. The details of our baryonic model are discussed in Appendix~\ref{app:baryons}. Briefly, we model the baryonic components of the Galaxy at the Solar cylindrical radius with 15 components (ten stellar and five gas components) as per Ref.~\cite{2015ApJ...814...13M}, with refinements from Refs.~\cite{2018PhRvL.121h1101S,2023arXiv230312838O}. Each component has an exponential or Gaussian suppression as a function of height $|z|$ off of the disk and an exponential in radial distance from the Galactic center -- note that all these baryonic components assume axial symmetry. The parameters of each model can be found in Table~\ref{table:baryonic_inputs}. Baryons dominate the mass density within $|z|\sim 0.5~$kpc of the disk. 

In Table~\ref{table:mass_density}, we report the total mass density, the baryonic density, and the inferred dark matter density calculated at the Solar location, using a single averaging kernel centered at the Sun.

\begin{figure*}
    \centering
    \includegraphics{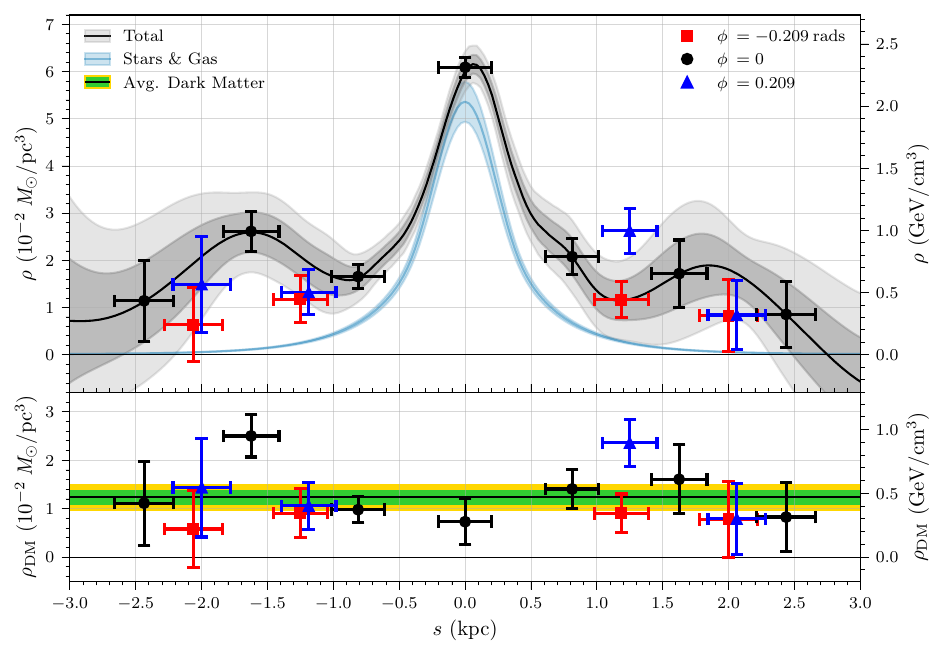}
    \caption{Top: Independent measurements of total mass density as a function of polar arclength $s$, for three values of $\phi$: $\phi = 0$ (black circles), $\phi = +0.209$ (blue triangles), and $\phi = -0.209$ (red squares). Points at nonzero $\phi$ have been offset for visibility. Horizontal error bars indicate the $1\sigma$ width of the kernel in the $s$-direction. The mass density at $\phi=0$ for $s$ values with overlapping kernels is shown in the black curve (note that due to kernel overlap these measurements are correlated at length scales of $\sim 0.4$~kpc). The baryonic mass density model is shown with the blue curve. Dark and light bands represent 1$\sigma$ and 2$\sigma$ uncertainties, respectively. Bottom: Independent measurements of dark matter mass density as a function of $s$ and $\phi$, obtained by subtracting the baryonic density in the top panel from the corresponding total mass density. The best-fit value of $\rho_{\mathrm{DM}}(r=r_\odot)=\localdensityGeV\pm\localdensityerrGeV$~$\mathrm{GeV/cm^3}$ is shown as a horizontal black line enclosed by green and yellow bands denoting the 1$\sigma$ and 2$\sigma$ uncertainties, respectively.
    }
    \label{fig:massdensity_z}
\end{figure*}

In Figure~\ref{fig:massdensity_z}, we show the total mass density $\rho$, the modeled baryonic mass, and the dark matter mass density as a function of arc length $s$ above and below the disk from the Solar location at $\phi = 0$. Due to the finite kernel size, we cannot calculate densities at the edge of the 4~kpc sphere centered on the Sun, and (as errors increase at larger distances) we instead show densities only up to $s=3$~kpc. 

The curve in Figure~\ref{fig:massdensity_z} is created from mass densities sampled with kernel centers more densely packed than the kernel size. As a result, neighboring density values are correlated. This smooths the resulting curve as a function of $s$, and statistical fluctuations in individual uncorrelated measurements appear as extended bumps. This effect is also the likely source of the slight off-set in the peak in the total mass density from $s=0$.

The mass densities at $15$ points at the Solar radius $r_\odot$ with independent, non-overlapping kernels\footnote{As MAFs are highly expressive models with a very large number of parameters, it is reasonable to assume that points in space separated at sufficiently large length scales are described by unique model parameters and are not correlated. For all practical purposes, if the averaging kernels of two mass density estimates do not overlap, the two estimates are independent measurements. \nopagebreak}
are indicated in Figure~\ref{fig:massdensity_z}. We use these $15$ independent measurements of the total mass density at $r_\odot$ to obtain $15$ independent estimates of the dark matter density at the Solar radius after subtracting the baryonic density profile. We find that these $15$ independent measurements of the dark matter density are consistent with a constant value ($\chi_\nu^2=1.38$), with a slight fluctuation below the average value at the Solar location and two more significant upwards fluctuations at $s\approx-1.6$~kpc and $s\approx1.25$~kpc. Assuming a spherically symmetric dark matter distribution, we report the average dark matter density at the Solar radius in Table~\ref{table:mass_density}. Note that the assumption of a spherically symmetric dark matter profile is an imposition of a symmetry assumption that has not been used elsewhere in this analysis so far.

\begin{figure*}
    \centering
    \resizebox{4.8in}{!}{
    \begin{tikzpicture}[every node/.style={inner sep=0,outer sep=0}]
        \node [anchor=south west] at (0,0) {\includegraphics{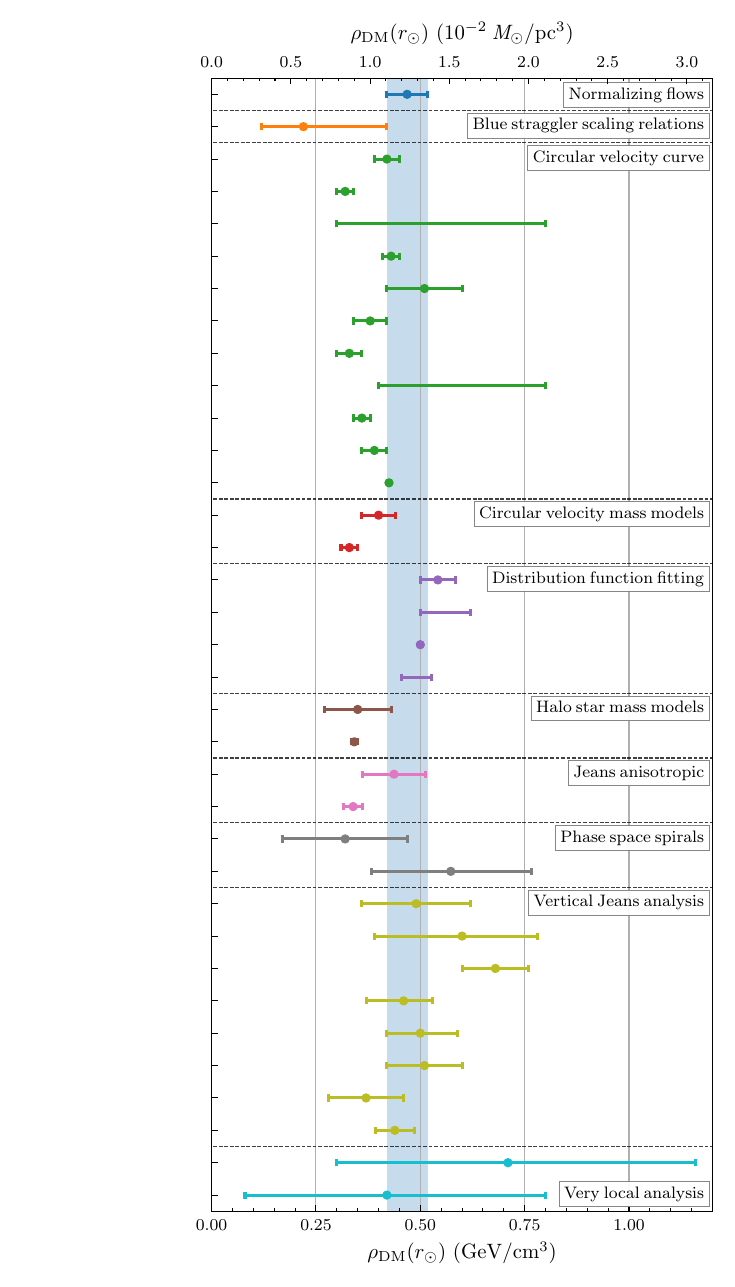}};
        \draw [white] (0in, 0) -- (0in, 8.57847 in);
        \node [draw=none, anchor=east, align=right] at (1.4in, 7.94182in) {
            \textbf{This work}
        };
        \node [draw=none, anchor=east, align=right] at ($ (1.4in,  7.94182in) - (0in, 0.21572in) $) {
            Casagrande, (2020) \cite{2020ApJ...896...26C}
        };
        \node [draw=none, anchor=east, align=right] at ($ (1.4in,  7.94182in) - 2*(0in, 0.21572in) $) {
            Pato, et al., (2015) \cite{2015JCAP...12..001P}
        };
        \node [draw=none, anchor=east, align=right] at ($ (1.4in,  7.94182in) - 3*(0in, 0.21572in) $) {
            Huang, et al., (2016) \cite{2016MNRAS.463.2623H}
        };
        \node [draw=none, anchor=east, align=right] at ($ (1.4in,  7.94182in) - 4*(0in, 0.21572in) $) {
            Benito, et al., (2019) \cite{2019JCAP...03..033B}
        };
        \node [draw=none, anchor=east, align=right] at ($ (1.4in,  7.94182in) - 5*(0in, 0.21572in) $) {
            Karukes, et al., (2019) \cite{2019JCAP...09..046K}
        };
        \node [draw=none, anchor=east, align=right] at ($ (1.4in,  7.94182in) - 6*(0in, 0.21572in) $) {
            Lin, et al., (2019) \cite{2019MNRAS.487.5679L}
        };
        \node [draw=none, anchor=east, align=right] at ($ (1.4in,  7.94182in) - 7*(0in, 0.21572in) $) {
            de Salas, et al., (2019) \cite{2019JCAP...10..037D}
        };
        \node [draw=none, anchor=east, align=right] at ($ (1.4in,  7.94182in) - 8*(0in, 0.21572in) $) {
            Ablimit, et al., (2020) \cite{2020ApJ...895L..12A}
        };
        \node [draw=none, anchor=east, align=right] at ($ (1.4in,  7.94182in) - 9*(0in, 0.21572in) $) {
            Benito, et al., (2020) \cite{2021PDU....3200826B}
        };
        \node [draw=none, anchor=east, align=right] at ($ (1.4in,  7.94182in) - 10*(0in, 0.21572in) $) {
            Sofue, (2020) \cite{2020Galax...8...37S}
        };
        \node [draw=none, anchor=east, align=right] at ($ (1.4in,  7.94182in) - 11*(0in, 0.21572in) $) {
            Zhou, et al., (2022) \cite{2022arXiv221210393Z}
        };
        \node [draw=none, anchor=east, align=right] at ($ (1.4in,  7.94182in) - 12*(0in, 0.21572in) $) {
            Ou, et al., (2023) \cite{2023arXiv230312838O}
        };
        \node [draw=none, anchor=east, align=right] at ($ (1.4in,  7.94182in) - 13*(0in, 0.21572in) $) {
            McMillan, (2017) \cite{2017MNRAS.465...76M}
        };
        \node [draw=none, anchor=east, align=right] at ($ (1.4in,  7.94182in) - 14*(0in, 0.21572in) $) {
            Cautun, et al., (2020) \cite{2020MNRAS.494.4291C}
        };
        \node [draw=none, anchor=east, align=right] at ($ (1.4in,  7.94182in) - 15*(0in, 0.21572in) $) {
            Bienyame, et al., (2014) \cite{2014A&A...571A..92B}
        };
        \node [draw=none, anchor=east, align=right] at ($ (1.4in,  7.94182in) - 16*(0in, 0.21572in) $) {
            Piffl, et al., (2014) \cite{2014MNRAS.445.3133P}
        };
        \node [draw=none, anchor=east, align=right] at ($ (1.4in,  7.94182in) - 17*(0in, 0.21572in) $) {
            Binney, et al., (2015) \cite{2015MNRAS.454.3653B}
        };
        \node [draw=none, anchor=east, align=right] at ($ (1.4in,  7.94182in) - 18*(0in, 0.21572in) $) {
            Cole, et al., (2017) \cite{2017MNRAS.465..798C}
        };
        \node [draw=none, anchor=east, align=right] at ($ (1.4in,  7.94182in) - 19*(0in, 0.21572in) $) {
            Wegg, et al., (2019) \cite{2019MNRAS.485.3296W}
        };
        \node [draw=none, anchor=east, align=right] at ($ (1.4in,  7.94182in) - 20*(0in, 0.21572in) $) {
            Hattori, et al., (2020) \cite{2021MNRAS.508.5468H}
        };
        \node [draw=none, anchor=east, align=right] at ($ (1.4in,  7.94182in) - 21*(0in, 0.21572in) $) {
            Nitschai, et al., (2020) \cite{2020MNRAS.494.6001N}
        };
        \node [draw=none, anchor=east, align=right] at ($ (1.4in,  7.94182in) - 22*(0in, 0.21572in) $) {
            Nitschai, et al., (2021) \cite{2021ApJ...916..112N}
        };
        \node [draw=none, anchor=east, align=right] at ($ (1.4in,  7.94182in) - 23*(0in, 0.21572in) $) {
            Widmark, et al., (2021) \cite{2021A&A...653A..86W}
        };
        \node [draw=none, anchor=east, align=right] at ($ (1.4in,  7.94182in) - 24*(0in, 0.21572in) $) {
            Guo, et al., (2022) \cite{2022ApJ...936..103G}
        };
        \node [draw=none, anchor=east, align=right] at ($ (1.4in,  7.94182in) - 25*(0in, 0.21572in) $) {
            McKee, et al., (2015) \cite{2015ApJ...814...13M}
        };
        \node [draw=none, anchor=east, align=right] at ($ (1.4in,  7.94182in) - 26*(0in, 0.21572in) $) {
            Xia, et al., (2016) \cite{2016MNRAS.458.3839X}
        };
        \node [draw=none, anchor=east, align=right] at ($ (1.4in,  7.94182in) - 27*(0in, 0.21572in) $) {
            Hagen, et al., (2018) \cite{2018A&A...615A..99H}
        };
        \node [draw=none, anchor=east, align=right] at ($ (1.4in,  7.94182in) - 28*(0in, 0.21572in) $) {
            Sivertsson, et al., (2018) \cite{2018MNRAS.478.1677S}
        };
        \node [draw=none, anchor=east, align=right] at ($ (1.4in,  7.94182in) - 29*(0in, 0.21572in) $) {
            Guo, et al., (2020) \cite{2020MNRAS.495.4828G}
        };
        \node [draw=none, anchor=east, align=right] at ($ (1.4in,  7.94182in) - 30*(0in, 0.21572in) $) {
            Salomon, et al., (2020) (North) \cite{2020A&A...643A..75S}
        };
        \node [draw=none, anchor=east, align=right] at ($ (1.4in,  7.94182in) - 31*(0in, 0.21572in) $) {
            Salomon, et al., (2020) (South) \cite{2020A&A...643A..75S}
        };
        \node [draw=none, anchor=east, align=right] at ($ (1.4in,  7.94182in) - 32*(0in, 0.21572in) $) {
            Wardana, et al., (2020) \cite{2020EPJWC.24004002W}
        };
        \node [draw=none, anchor=east, align=right] at ($ (1.4in,  7.94182in) - 33*(0in, 0.21572in) $) {
            Schutz, et al., (2018) \cite{2018PhRvL.121h1101S}
        };
        \node [draw=none, anchor=east, align=right] at ($ (1.4in,  7.94182in) - 34*(0in, 0.21572in) $) {
            Buch, et al., (2019) \cite{2019JCAP...04..026B}
        };
    \end{tikzpicture}
    }
    \caption{Our averaged measurement of the dark matter density at the Solar radius (top line), compared to recent measurements \cite{2021ApJ...916..112N,2020ApJ...896...26C,2015JCAP...12..001P,2016MNRAS.463.2623H,2019JCAP...03..033B,2019JCAP...09..046K,2019MNRAS.487.5679L,2019JCAP...10..037D,2020ApJ...895L..12A,2021PDU....3200826B,2020Galax...8...37S,2022arXiv221210393Z,2023arXiv230312838O,2017MNRAS.465...76M,2020MNRAS.494.4291C,2014A&A...571A..92B,2015MNRAS.451..639P,2014MNRAS.445.3133P,2015MNRAS.454.3653B,2017MNRAS.465..798C,2019MNRAS.485.3296W,2021MNRAS.508.5468H,2020MNRAS.494.6001N,2021A&A...653A..86W,2022ApJ...936..103G,2015ApJ...814...13M,2016MNRAS.458.3839X,2018A&A...615A..99H,2018MNRAS.478.1677S,2020MNRAS.495.4828G,2020A&A...643A..75S,2020EPJWC.24004002W,2018PhRvL.121h1101S,2019JCAP...04..026B} of the density of dark matter at or near the Solar location $\rho_{\rm {DM},\odot}$. See \cite{2021RPPh...84j4901D} for a detailed review of most of these measurements and their techniques. Our measurement of $\overline{\rho}_{\rm {DM}}(r=r_\odot)=\localdensity\pm\localdensityerr\times10^{-2}\,M_\odot/{\rm pc}^3= \localdensityGeV\pm\localdensityerrGeV\,{\rm GeV/cm}^3$ is consistent with the existing literature of measurements of $\rho_{\rm {DM},\odot}$.
    }
    \label{fig:darkmatter_densities}
\end{figure*}

In Figure~\ref{fig:darkmatter_densities}, we compare our result for $\overline{\rho}_{\rm DM}(r=r_\odot)$ with a set of recent measurements from the last decade that use a variety of methods. As can be seen, our result is consistent with the previous literature, with competitive and comprehensive error bars. Our errors contain both statistical and 
measurement uncertainties in the calculation of the total mass density, as well as the reported uncertainties for our baryonic density model described in Appendix~\ref{app:baryons}.

\begin{table}[t]
\centering
\begin{tabular}{cccc}
    \toprule
    Density &
    $(10^{-2}\:M_\odot/{\rm pc}^3)$ &
    $(\rm GeV/cm^3)$ &
    $\chi_\nu^2$
    \\
    \midrule
    $\rho_\odot$ &
    $6.17 \pm 0.20$ & 
    $2.34 \pm 0.08\w$ &
    
    \\
    $\rho_{b,\odot}$ & 
    $5.34 \pm 0.42$ & 
    $2.03 \pm 0.16\w$  &
    
    \\
    $\rho_{\rm DM,\odot}$ & 
    $0.83 \pm 0.47$ & 
    $0.32 \pm 0.18\w$ &
    
    \\
    \midrule
    $\overline{\rho}_{\rm DM}(r=r_\odot)$ &
    $\localdensity \pm \localdensityerr$ &
    $\localdensityGeV \pm \localdensityerrGeV\w$ &
    $1.38$
    \\
    \bottomrule
\end{tabular}
\caption{MAF-estimated densities at the Solar location or averaged at the Solar radius $r_\odot$.
The dark matter density is the difference between the total mass density $\rho_\odot$ and the baryonic mass density (obtained from the model described in Appendix~\ref{app:baryons}).
The averaged mass density at $r=r_\odot$ is the weighted average of the dark matter mass density evaluated at 15 independent points at $r=r_\odot$ 
(as in Figure~\ref{fig:massdensity_z}) and subtracting the baryonic mass density in that region.
}
\label{table:mass_density}
\end{table}

\begin{figure*}
   \centering   
   \includegraphics[width=\textwidth]{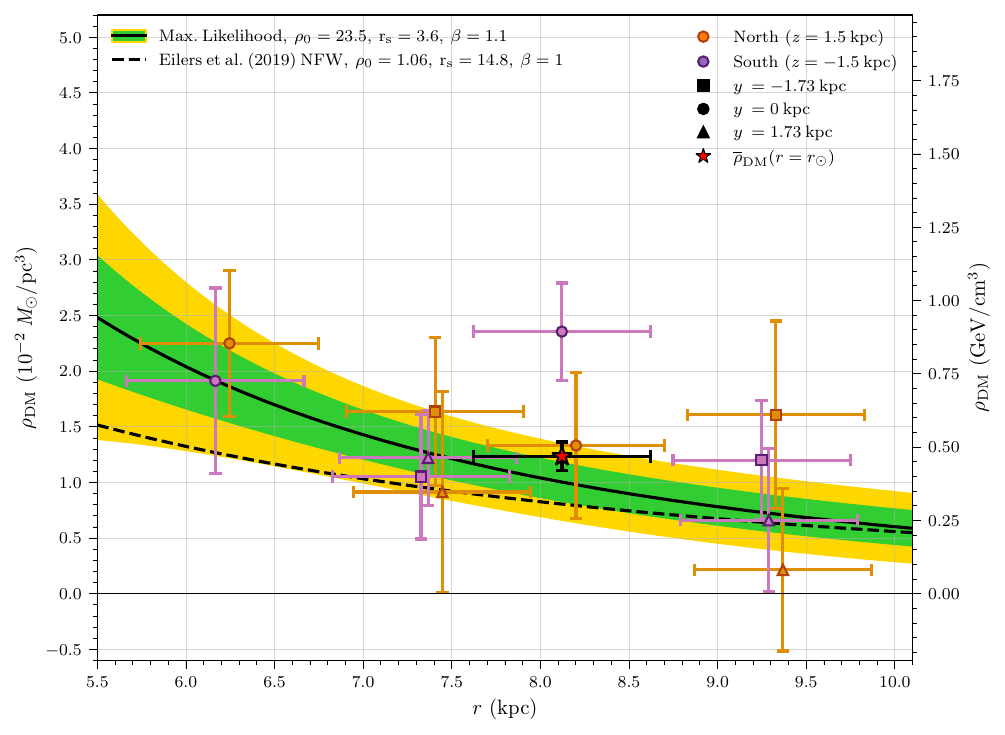}
   \caption{
   Dark matter mass densities measured at 12 independent points as a function of Galactic spherical radius $r$. Orange points were evaluated at $z=+1.5$~kpc, purple points were evaluated at $z=-1.5$~kpc. The average dark matter density at the Solar radius $r_\odot$ is shown as a red star. 
   The maximum likelihood fit to a generalized NFW profile is shown in a solid black line, with 1$\sigma$ and $2\sigma$ variance across the posterior distribution of models explored in Figure~\ref{fig:gnfw_fit} shown as green and yellow error bands, respectively. A recent fit \cite{2019ApJ...871..120E} to a standard NFW profile to the rotation curve of the Milky Way is shown as a dashed black line.
   }
   \label{fig:radial_profile_and_gnfw_fit}
\end{figure*}

If we again assume spherical symmetry, we can further investigate the dependence of the dark matter mass density on spherical radius $r$. We show in Figure~\ref{fig:radial_profile_and_gnfw_fit} the dark matter mass densities evaluated at various Galactocentric radii $r$, along $z=\pm 1.5$~kpc off-sets from the disk (and at different values of $\phi$). 
Considering independent measurements over a range of $r$ values within the 4~kpc observational sphere, we fit the measured $\rho_{\rm DM}$ as a function of $r$ to the generalized NFW profile
\begin{equation}
    \rho_{\rm DM}(r) = \frac{\rho_0}{\left(\frac{r}{r_s}\right)^\beta\left(1+\frac{r}{r_s}\right)^{3-\beta}}, \label{eq:nfw_form}
\end{equation}
with free parameters $\rho_0$, $r_s$, and $\beta$ (in the standard NFW, $\beta=1$). In performing our fit, we adopted a truncated Gaussian prior for $\beta$ centered at $2$ with a width of 2. We adopt a flat prior for $r_s$ within $[0,20]$, and a truncated Gaussian prior centered at $40\times 10^{-2}M_\odot/{\rm pc}^3$ with a width of $80\times 10^{-2}M_\odot/{\rm pc}^3$ for $\rho_0$ in the range $[0,200]\times 10^{-2}\,M_\odot/{\rm pc}^3$. This choice of priors restricts the model space to a physically realistic domain: the Gaussian prior for $\rho_0$ prevents arbitrarily large density parameters, leading to extremely small values of the scale radius $r_s$. 

\begin{figure}
   \centering
   \includegraphics[width=0.6\columnwidth]{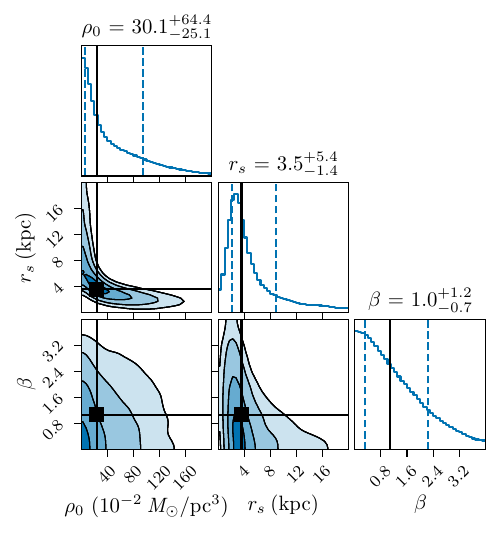}
   \caption{
   The posterior distribution for a fit to a generalized NFW profile Eq.~\eqref{eq:nfw_form}. The median values of each parameter are shown above the marginalized 1D histograms, with the $16^{\rm th}$ and $84^{\rm th}$ percentile values shown as error bounds.  The maximum likelihood model is shown in black. 
   }
   \label{fig:gnfw_fit}
\end{figure}

We show the posterior distribution for $\rho_0$, $r_s$, and $\beta$ in Figure~\ref{fig:gnfw_fit} with median values and 16-th and 84-th percentile uncertainties of $\rho_0=30.1_{-25.1}^{+64.4}\times 10^{-2}M_\odot/{\rm pc}^3$, $r_s=3.5_{-1.4}^{+5.4}$~kpc, and $\beta = 1.0_{-0.7}^{+1.2}$. The best fit model is $\rho_0=23.5\times 10^{-2}M_\odot/{\rm pc}^3$, $r_s=3.6$~kpc, and $\beta = 1.1$. This is broadly in agreement with other recent fits to the dark matter density profile (e.g., Refs.~\cite{2019ApJ...871..120E,2023arXiv230312838O}). We plot our best-fit profile in Figure~\ref{fig:radial_profile_and_gnfw_fit}. We note that -- given the range of complete data available from \Gaia{}, measurement errors, and dust extinction -- our dataset does not extend to the low-$r$ regime, and so does not yet provide significant discriminating power between different models of dark matter density profiles.

\section{Discussion} \label{sec:discussion}

Using normalizing flows to model the phase space density of bright stars within \Gaia{} DR3, we have -- for the first time -- measured the gravitational acceleration and mass density within the local volume of the Milky Way without assumption of functional form or symmetry. The resulting acceleration and mass density maps across a three dimensional volume around the Sun provide a unique window into Galactic structure and dynamics.

We report the acceleration at the Solar location, with $a_r = (\aR\pm \aRerr)\times 10^{-10}$~m/s$^2$. This is almost within $2\sigma$ of recent measurements of the acceleration using quasars \cite{gaia_edr3_accel}.
Deviations from axisymmetry and the assumptions of equilibrium are found when comparing the measurements in the Galactic North and South. Though at this time the statistical and systematic errors on our MAF-derived accelerations are too large to draw confident conclusions about departures from equilibrium, our method allows us to test assumptions of equilibrium and symmetry in a way that has not been previously possible.

Without assuming any symmetries, our measurement of dark matter at the Solar location is $\rho_{\rm DM,\odot} = 0.32\pm0.18$~GeV/cm$^3$. This large uncertainty can be significantly reduced with the further imposition of spherical symmetry -- allowing the averaging of measurements at different locations. Under this assumption, we find a dark matter density at the Solar radius of $\localdensityGeV\pm \localdensityerrGeV$~GeV/cm$^3$, in agreement with previous measurements using a variety of other techniques. Recall that our density measurements encompass a wide range of possible dark matter density profiles due to the expressivity of the MAF; this variation is fully incorporated into our error budget.

We fit sampled points of our dark matter density profile to a generalized NFW profile, and find a preference for a relatively small scale radius $r_s$ consistent with a recent measurement of the Milky Way's rotation curve \cite{2023arXiv230312838O}. However, our current range of reliable data does not yet extend deep into the central region of the Galaxy, and so our statistical preference for this value of $r_s$ is not high.

These machine-learning assisted measurements of Galactic acceleration and mass density are expected to improve significantly in the near future. \Gaia{} DR4 and DR5 are expected to expand the number of stars with full six-dimensional kinematics by a factor of three \cite{2022arXiv220605902K}, with an associated decrease in statistical errors in our determination of the phase space density. Measurement errors are likewise expected to improve by a factor of two \cite{gaia_science_performance_2022, 2022arXiv220800211G}, and improvements in our analysis technique to correct for the bias introduced by these errors
are possible.

Dust extinction is a major limitation in applying our algorithm to regions close to the disk or toward the Galactic center. An improved understanding of the effects of dust on the measured \Gaia{} stellar features which are uniform across the sky will allow greater accuracy in our measurements of phase space, acceleration, and mass density. Indeed, normalizing flows may play a role in data-driven modelling of dust extinction, which we will pursue in future work. 

Overall, we can expect improvements in architecture, analysis, data quantity, and data quality to allow great advances over these first results. In addition to greatly improved precision and better constraints on the density profile, future analyses based on these techniques may be able to directly probe the departures from equilibrium within the Milky Way, especially when combined with other measurements of local acceleration, such as those based on quasars \cite{gaia_edr3_accel}, pulsars \cite{2021ApJ...907L..26C}, or binary systems \cite{2022ApJ...928L..17C}.

\section*{Acknowledgements}

We thank Adrian Price-Whelan and Mitchell Weikert for helpful discussions. This work was supported by the DOE under Award Number DOE-SC0010008. 
The work of SHL was also partly supported by IBS under the project code, IBS-R018-D1. 
This work was also performed in part at Aspen Center for Physics, which is supported by National Science Foundation grant PHY-2210452.
This work has made use of data from the European Space Agency (ESA) mission \Gaia{} (\url{https://www.cosmos.esa.int/gaia}), processed by the \Gaia{} Data Processing and Analysis Consortium (DPAC, \url{https://www.cosmos.esa.int/web/gaia/dpac/consortium}). Funding for the DPAC has been provided by national institutions, in particular the institutions participating in the \Gaia{} Multilateral Agreement. We thank the \Gaia{} Project Scientist Support Team and DPAC for their work in development and maintenance of the \textsc{PyGaia} code. 
The authors acknowledge the Office of Advanced Research Computing (OARC) at Rutgers, The State University of New Jersey for providing access to the Amarel cluster and associated research computing resources that have contributed to the results reported here. URL: \url{https://oarc.rutgers.edu}

\section*{Data Availability}
This work uses the \Gaia{} DR3 dataset hosted in FlatHUB repository, at \url{{https://flathub.flatironinstitute.org/gaiadr3}}. 
The processed datasets underlying this article will be shared on reasonable request to the corresponding author.

\appendix

\section{Baryon Mass Model} \label{app:baryons}
In order to estimate the local density field of dark matter $\rho_{\rm DM}(\vec{x})$ given the total density field $\rho(\vec{x})$, we must estimate the local distribution of baryonic mass density $\rho_{b}(\vec{x})$ in the Milky Way. We follow Refs.~\cite{2018PhRvL.121h1101S,2022MNRAS.511.1977S,2021A&A...653A..86W,2021ApJ...907L..26C,2021A&A...646A..67W,2019JCAP...04..026B,2018A&A...615A..99H,2016ApJ...824..116K} and base our model for $\rho_{b}(\vec{x})$ in the Solar neighborhood on the work of Ref.~\cite{2015ApJ...814...13M} (hereafter referred to as the McKee model), an extensive compilation of estimates of the surface mass densities of gas, stars, and compact objects in the Solar neighborhood.

The McKee mass model is broken into 15 components: five types of gas, seven stellar populations, and three populations of compact objects. Each component is described in Table~\ref{table:baryonic_inputs}. The McKee model draws on a collection of pre-\Gaia{} star counts and gas surveys. Recent updates using \Gaia{} data \cite{2017MNRAS.470.1360B,2023arXiv230312838O} have not substantially altered the model. The stellar bulge and halo do not significantly contribute to the mass density in the Solar neighborhood, and so they are not independently modelled. However, this model does include halo stars within the disk. 

The McKee model characterizes the surface densities $\Sigma_{0,i}$ and effective scale heights $h_{z,i}$, as well as the functional form for the number density as a function of $z$, for each component. Assuming direct proportionality between number and mass density -- i.e., assuming no chemical evolution of each component as a function of $z$ -- we can model each mass density in the same way as the number density. All but 12 components were fit to an exponential mass density profile
\begin{equation}
\rho_{\rm exp,i}(\vec{x}) = \frac{\Sigma_{0,i}}{h_i}e^{-|z|/h_i}e^{-(R-r_\odot)/h_R}.
\end{equation}
The remaining three components ($\mathrm{H_2}$, $\mathrm{HI}_{\rm CNM}$, and $\mathrm{HI}_\mathrm{WNM,1}$) were fit to the following Gaussian mass density profile
\begin{equation}
\rho_{\rm gauss,i}(\vec{x}) = \frac{\Sigma_{0,i}}{\sqrt{\pi}h_i}e^{-|z|^2/h_i^2}e^{-(R-r_\odot)/h_R}.
\end{equation}
We supplement each component of the McKee model with a radial scale length $h_{R,i}$, informed by the baryonic model used in Ref.~\cite{2023arXiv230312838O}. Stellar populations were assigned a scale radius of 2.35~kpc, all $\mathrm{HI}$ gasses were assigned a large scale radius of $18.24$~kpc, and $\mathrm{H_2}$ gas was assigned $h_R=2.57$~kpc. $\mathrm{HII}$ gas was assigned an arbitrary scale radius of 2.5~kpc. Due to its overall small contribution to the surface density, uncertainty in this scale length does not have a significant effect on the mass density. It should also be emphasized that the precise details of the baryonic radial profile are insignificant for $|z|>500$~pc and for all $R$ within our observational window.

Values and uncertainties (when available) for $\Sigma_{0,i}$, $h_{z,i}$, $h_{R,i}$, and the corresponding $\rho_{0,i} \equiv \rho_i(z=0)$ for all 15 components are given in Table~\ref{table:baryonic_inputs}. We follow Ref.~\cite{2018PhRvL.121h1101S} in assigning $10\%$ uncertainties to any unreported surface density errors in the McKee model. In total, we expect approximately $8\%$ uncertainty in the baryonic mass density at the Solar location $\rho_\mathrm{b,\odot}$. We do not follow Ref.~\cite{2015ApJ...814...13M} in inflating this uncertainty to $15\%$, although we agree that the systematic uncertainties in the original error model are likely underestimated.

\begin{table*}[th]
\centering
\begin{tabular}{lccccc}
\toprule
\multicolumn{1}{c}{Population} & $\Sigma_{0,i}$ ($\textit{M}_\odot/\mathrm{pc^2}$) & $h_{z,i}$ (kpc) & $h_{R,i}$ (kpc) & $\rho_{0,i}$ ($10^{-2}\:\textit{M}_\odot/\mathrm{pc^3}$) & \multicolumn{1}{c}{Refs.} \\
\midrule
\bf Gas & $13.66\pm1.32$ & & & $\w4.064\pm0.577$ & \\
$\text{HI}_\text{CNM}$    & $\w6.21\pm1.24$  & \w0.127 & 18.24 & $\w2.759\pm0.552$ & \cite{2015ApJ...814...13M,2023arXiv230312838O}\\
H$_2$                     & $\w1.00\pm0.3\w$  & \w0.105 & \w2.57 & $\w0.537\pm0.161$ & \cite{2015ApJ...814...13M,2023arXiv230312838O}\\
$\text{HI}_\text{WNM,1}$   & $\w2.51\pm0.25$  & \w0.318 & 18.24 & $\w0.445\pm0.045$ & \cite{2015ApJ...814...13M,2023arXiv230312838O}\\
$\text{HI}_\text{WNM,2}$   & $\w2.14\pm0.21$  & \w0.403 & 18.24 & $\w0.266\pm0.027$ & \cite{2015ApJ...814...13M,2023arXiv230312838O}\\
HII                        & $\w1.80\pm0.1\w$  & \w1.590 & \w2.50 & $\w0.057\pm0.003$ & \cite{2015ApJ...814...13M}\\
\midrule
\bf Stars & $28.2\w\pm2.42$ & & & $\w3.730\pm0.303$ & \\
M Dwarves                  & $17.30\pm2.3\w$   & \w0.400 & \w2.35 & $\w2.163\pm0.287$ & \cite{2015ApJ...814...13M,2023arXiv230312838O} \\
5 $<$ $M_V$ $<$ 8          & $\w5.80\pm0.58$  & \w0.400 & \w2.35 & $\w0.725\pm0.072$ & \cite{2015ApJ...814...13M,2023arXiv230312838O}\\
4 $<$ $M_V$ $<$ 5          & $\w2.20\pm0.22$  & \w0.384 & \w2.35 & $\w0.286\pm0.029$ & \cite{2015ApJ...814...13M,2023arXiv230312838O}\\
$M_V$ $<$ 3                & $\w0.50\pm0.05$  & \w0.140 & \w2.35 & $\w0.179\pm0.018$ & \cite{2015ApJ...814...13M,2023arXiv230312838O}\\
3 $<$ $M_V$ $<$ 4          & $\w0.80\pm0.08$  & \w0.236 & \w2.35 & $\w0.169\pm0.017$ & \cite{2015ApJ...814...13M,2023arXiv230312838O}\\
Brown Dwarves              & $\w1.20\pm0.40$  & \w0.400 & \w2.35 & $\w0.150\pm0.050$ & \cite{2015ApJ...814...13M,2023arXiv230312838O}\\
Giants                     & $\w0.40\pm0.04$  & \w0.344 & \w2.35 & $\w0.058\pm0.006$ & \cite{2015ApJ...814...13M,2023arXiv230312838O}\\
\midrule
\bf Compact Objects & $\w5.2\w\pm0.80$ & & & $\w0.607\pm0.093$ & \\
White Dwarves              & $\w4.90\pm0.80$  & \w0.430 & \w2.35 & $\w0.570\pm0.093$ & \cite{2015ApJ...814...13M,2023arXiv230312838O}\\
Neutron Stars              & $\w0.20\pm0.05$  & \w0.400 & \w2.35 & $\w0.025\pm0.006$ & \cite{2015ApJ...814...13M,2023arXiv230312838O}\\
Black Holes                & $\w0.10\pm0.02$  & \w0.400 & \w2.35 & $\w0.013\pm0.003$ & \cite{2015ApJ...814...13M,2023arXiv230312838O}\\
\midrule
\bf Total & $47.06\pm2.87$ & & & $\w8.4\w\w\pm0.66$ & \\
\bottomrule
\end{tabular}
\label{table:baryonic_inputs}
\caption{All parameters of the baryonic mass model used in this work, as well as their respective references. $\Sigma_{0,i}$ is the surface density of each baryonic component at the Solar radius $R=r_\odot=8.122$~kpc. $h_{z,i}$ is the scale height, indicating how far above and below the midplane each component extends. As in Ref.~\cite{2015ApJ...814...13M}, some values of $h_{z,i}$ are a weighted average of two scale heights, representing an ``effective'' scale height. 
$\rho_{0,i}=\rho(z=0)_i$, where $\rho(z=0)_i$ is the volume mass density of each component in the midplane. $\rho(z=0)_i$ is computed from $\Sigma_{0,i}$ and $h_{z,i}$ via $\rho(z)_i=(1/2)\partial\Sigma(z)_i/\partial z$ and by assuming a form for $\Sigma(z)$. For an exponential mass profile, $\Sigma(z)_i=\Sigma_{0,i}(1-\exp(-|z|/h_{z,i}))$. $h_{R,i}$ is the exponential scale radius of each component, capturing the first-order radial behavior of baryonic mass density (ignoring detailed features such as spiral arms or clouds).}
\end{table*}

As discussed in Ref.~\cite{2021RPPh...84j4901D}, de-projecting the McKee model out of the plane into a volume density $\rho_b(z)$ comes with systematic uncertainties. Based on comparisons to the \textsc{MWPotential2014} Milky Way mass model implemented in the \textsc{galpy} library, our de-projection of the McKee model does not deviate significantly from other standard baryonic mass distributions. Additionally, these systematic uncertainties become subdominant to our other measurement and statistical uncertainties of $\rho(z)$ for $|z|>500$~pc, where $\rho_{b}$ is greatly sub-dominant to $\rho_{\rm DM}$. Only the estimate for the dark matter density at the Solar location $\rho_{\rm DM,\odot}$ is significantly affected by our choice and interpretation of the McKee model, as the Solar System is located near the midplane.

Finally, when evaluating the baryonic mass density at a particular point, we convolve our estimate for $\rho_\mathrm{b}(\vec{x})$ over the same quasi-random Gaussian kernel used to estimate $\rho(\vec{x})$. The convolved baryonic mass density profile $\rho_\mathrm{b}(\vec{x})*K$ is comparable to $\rho_\mathrm{b}(\vec{x})$ everywhere except for the disk, where the peak at $z=0$ is widened and shortened due to this convolution. As a result, if the vertical profile falls off too quickly with $z$, $\rho_\mathrm{b}(z=0)*K$ will be underestimated. 
This introduces systematic uncertainty in the estimate of $\rho_\mathrm{b}*K$ for $|z|\lesssim500$~pc in the disk, compared to $|z|\gtrsim500$~pc in the more robust halo region.

\section{Effect of Parallax Error Selection}
\label{app:parallax_err_sel}

As distant stars tend to have larger relative parallax errors, selecting stars with small relative parallax errors for the dataset quality control may introduce bias in the number density estimation.
This appendix qualitatively assesses the significance of bias due to the parallax error selection, $\sigma_\varpi / \varpi < 1/3$.

\begin{figure}[t]
    \centering
    \includegraphics[width=0.49\linewidth]{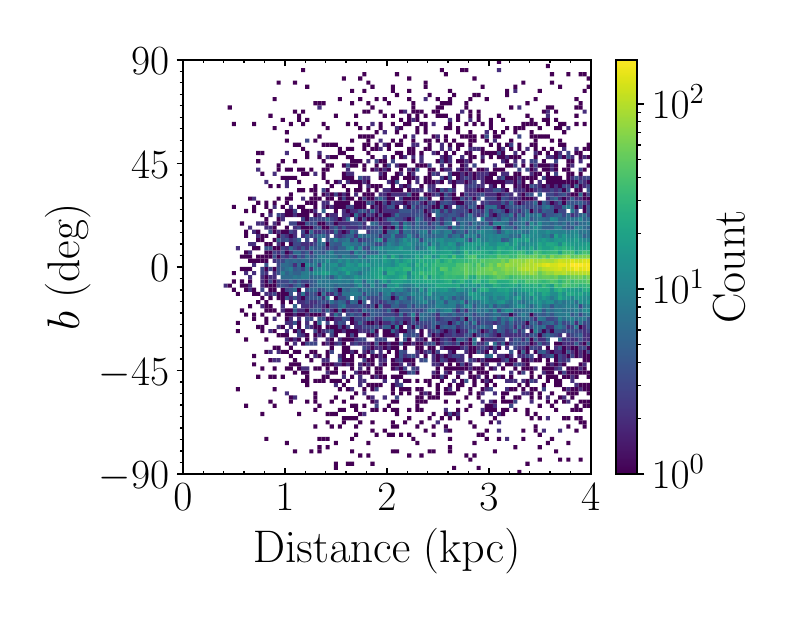}
    \includegraphics[width=0.49\linewidth]{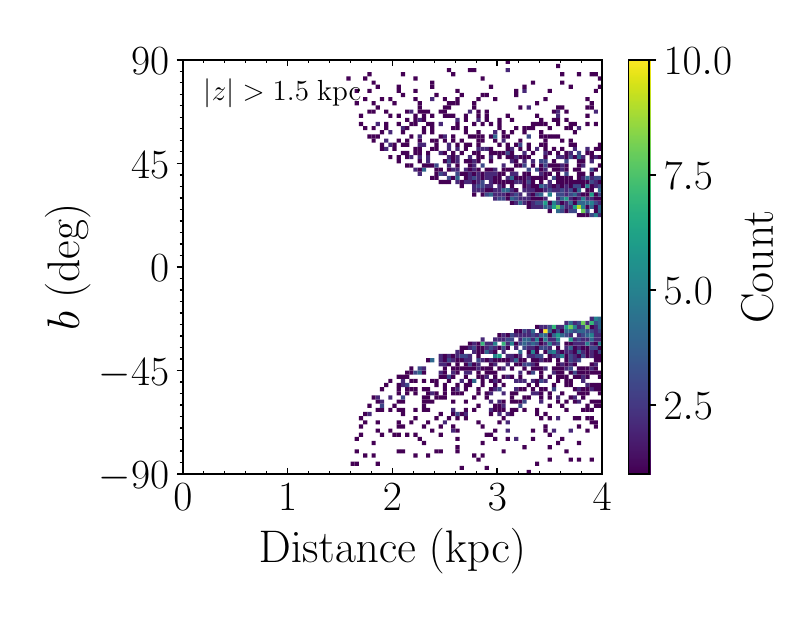}
    \caption{(left) Distance vs. galactic latitude histogram of the stars removed by parallax error selection and passing all the other criteria. (right) The same histogram as the left but with an additional criterion, $|z|>1.5$ kpc, to focus on the analysis region away from the disk.}
    \label{fig:d-b.parallax_error}
\end{figure}

The distribution of stars removed solely due to the selection criterion is shown in Figure~\ref{fig:d-b.parallax_error}. 
The high parallax error stars are either far away or close to the Milky Way's disk due to difficulties in measurement.
The distance vs. galactic latitude histogram on the left histogram of Figure~\ref{fig:d-b.parallax_error} clearly shows this tendency.
We can see two regions free from the parallax error selection effect. 
One is the nearby region with a distance of less than 0.5 to 1.0 kpc, as the parallax of a star here is generally large.
The other is the direction perpendicular to the disk, $|b|\sim90^{\circ}$.
This region is less crowded and dust-free; hence, the parallax of stars here can be measured more precisely than that of the stars close to the disk.
Since the analysis performed in this paper primarily focuses on these regions, we are mainly free from bias due to the bulk of the discarded stars.

However, the tail of the distribution extends to high galactic latitudes, potentially introducing bias in the analysis region away from the disk ($|z| > 1.5$ kpc), which is illustrated as a region covered by orange and purple lines in Figure~\ref{fig:coordinate_systems}.
The right histogram of Figure~\ref{fig:d-b.parallax_error} shows the distribution of the discarded stars with $|z| > 1.5$ kpc.
As distance increases along a line with a fixed $z$ value, the galactic latitude decreases, and the number of discarded stars increases.

\begin{figure}[t]
    \centering
    \includegraphics[width=0.75\linewidth]{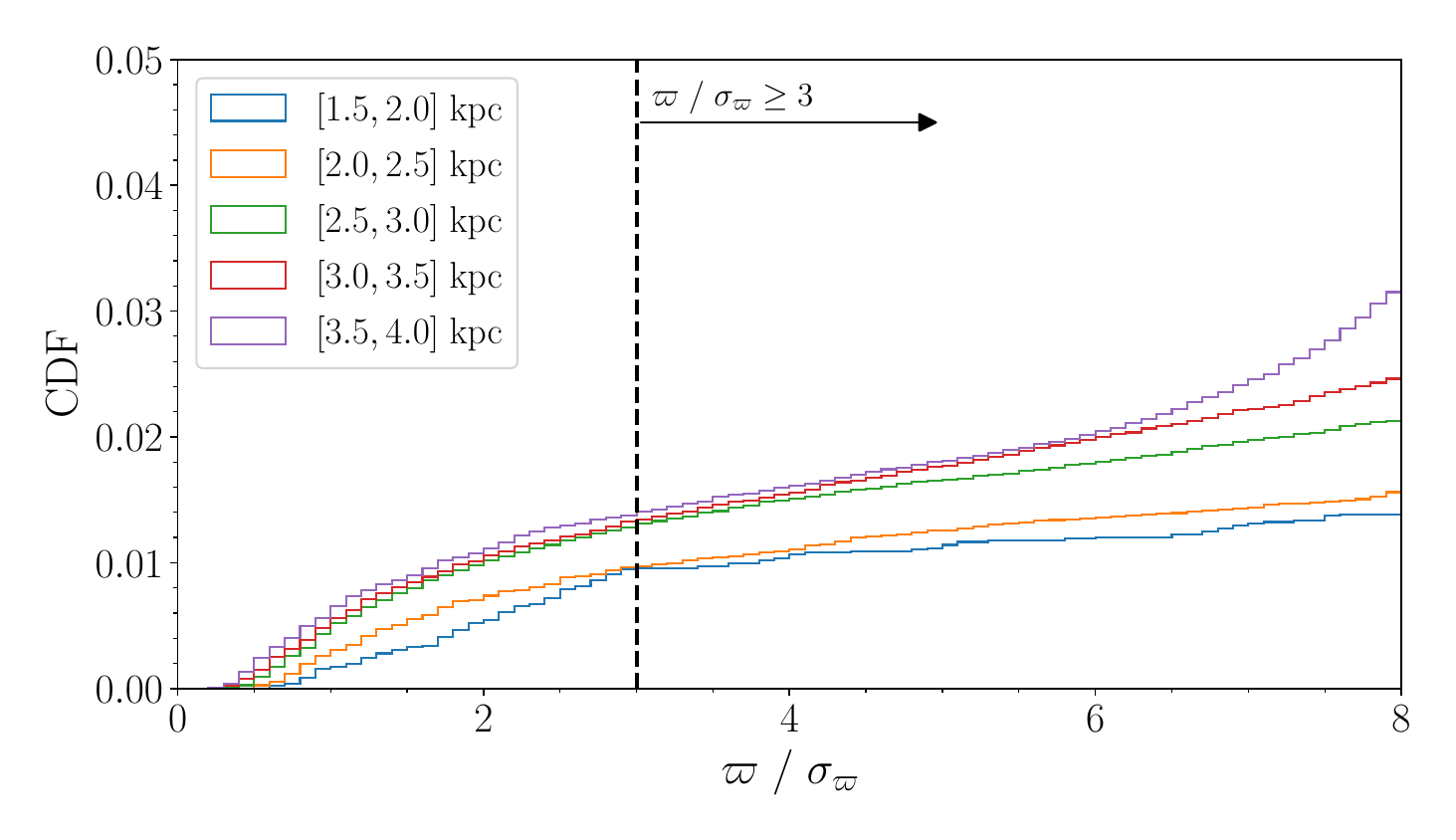}
    \caption{Cumulative histogram of parallax over error ($\varpi/\sigma_\varpi$) for each bin of distance. We only consider stars with $|z| > 1.5$ kpc and passing all the other selection criteria except the parallax error selection. Note that the cumulative distribution function (CDF) value corresponds to the discarded fraction in Table \ref{tab:discarded} given $\varpi/\sigma_\varpi$ selection threshold. }
    \label{fig:cdf_parallax_over_error}
\end{figure}

Table~\ref{tab:discarded} shows the fraction of the stars discarded by the parallax selection among the stars that passed all the other selection criteria.
The ratio is 0.95\% at a distance of 1.5 kpc and increases to 1.38\% at 3.5 kpc.
The cumulative histogram of the parallax over error $(\varpi/\sigma_\varpi)$ in Figure~\ref{fig:cdf_parallax_over_error} shows a more detailed response of these fractions over the change of the threshold value.
Consequently, there is a potential bias of at most 1.5\% in number density estimation due to this parallax error selection.
The increase in this fraction with distance may introduce a small bias in the density derivative estimation needed for solving the Boltzmann equation.

\begin{figure*}[t]
    \begin{center}
        \includegraphics[width=0.31\textwidth]{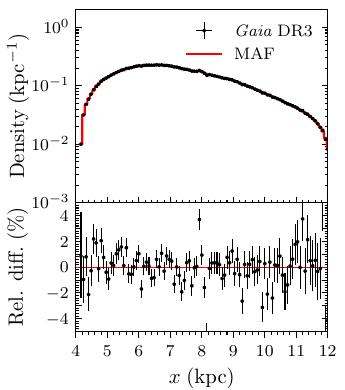}
        \includegraphics[width=0.31\textwidth]{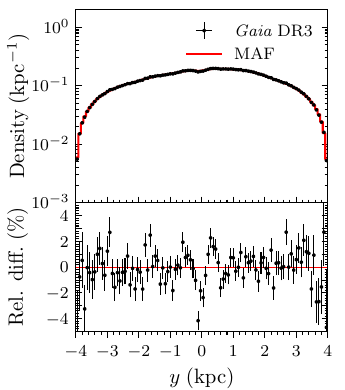}
        \includegraphics[width=0.31\textwidth]{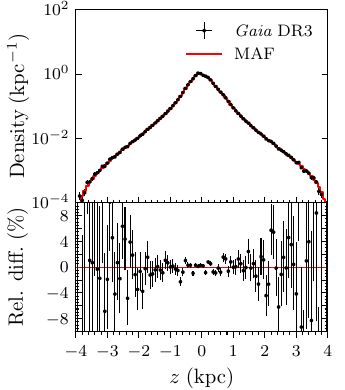}
    \end{center}
    \caption{
    Normalized histograms of position components for selected stars in \Gaia{} DR3 (downsampled to 20\% of the original size).
    The red lines are the histograms for synthetic stars sampled from the normalizing flows.
    The error bars are the $1\sigma$ statistical uncertainty.
    Below the main plots, we show the relative difference between \Gaia{} and MAF, i.e., the difference between \Gaia{} and MAF histograms divided by the MAF histogram. 
    }
    \label{fig:flow_density_rel_err}
\end{figure*}

Nevertheless, the fraction is smaller than the statistical uncertainty of MAF.
Figure~\ref{fig:flow_density_rel_err} again shows the one-dimensional histograms of the position components for the selected stars, and each figure's bottom panel shows the relative difference between \Gaia{} and MAF histograms. 
We can see that the difference is about 2\% in the Solar neighborhood and increases to about 4\% at the boundary of the fiducial volume.
Since this statistical variation is larger than the discarded fractions overall, we may safely ignore the effect of bias for the analysis within this paper.

\begin{table}[t]
\begin{center}
\begin{tabular}{cccccc}
\toprule
   Distance (kpc) & [1.5, 2.0] & [2.0, 2.5] & [2.5, 3.0] & [3.0, 3.5] & [3.5, 4.0] \\
\midrule
Discarded fraction (\%)  & 0.95 & 0.96 & 1.28 & 1.33 & 1.38 \\
\bottomrule
\end{tabular}
\end{center}
\caption{Fraction of removed stars by the parallax error selection for each bin of the distance. We only consider stars with $|z| > 1.5$ kpc and passing all the other selection criteria except the parallax error selection.}
\label{tab:discarded}
\end{table}

\bibliographystyle{JHEP}
\bibliography{density}

\end{document}